\documentclass{tlp1}

\usepackage{graphicx}

\newtheorem{notation}[subsubsection]{Notation}
\newtheorem{example}[subsubsection]{Example}
\newtheorem{corollaire}[subsubsection]{Corollary}
\newtheorem{definition1}[subsubsection]{Definition}
\newtheorem{propriete}[subsubsection]{Property}
\newtheorem{remarque}[subsubsection]{Remark}
\newtheorem{theoreme}[subsubsection]{Theorem}

\newcommand{\vrai}{\mathit{true}}
\newcommand{\faux}{\mathit{false}}
\newcommand{\Ff}{ F}
\newcommand{\Vv}{ V}

  \title[Theory and Practice of Logic Programming]
        {Decomposable Theories }

  \author[K. Djelloul]
         {KHALIL DJELLOUL \\Parc scientifique et technologique de
         Luminy\\
         Laboratoire d'Informatique Fondamentale de Marseille. \\163 avenue de Luminy - Case 901, 13288 Marseille, cedex 9.
         France\\
         \email{khalil.djelloul@lif.univ-mrs.fr}}

\begin{document}

\label{firstpage}

\maketitle

\noindent {\bfseries Note: } 
 This article will be published in Theory and Practice of Logic Programming (TPLP). \copyright Cambridge University Press. $\\[2mm]$

\begin{abstract}
We present in this paper a general algorithm for solving
first-order formulas in particular theories called
\emph{decomposable theories}. First of all, using special
quantifiers, we give a formal characterization  of 
 decomposable theories and show some of their properties. Then, we
present a general algorithm for solving first-order formulas in
 any decomposable theory $T$. The algorithm is given in the form of five rewriting rules.
It transforms a first-order formula $\varphi$, which can possibly
contain free variables, into a conjunction $\phi$ of solved formulas easily
transformable into a Boolean combination of existentially
quantified conjunctions of atomic formulas. In particular, if
$\varphi$ has no free variables then $\phi$ is either the formula
$\vrai$ or $\neg\vrai$. The correctness of our algorithm proves
the completeness of the decomposable theories.

Finally, we show that the theory ${\cal T}$ of finite or infinite
trees is a decomposable theory and give some benchmarks realized
by an implementation of our algorithm, solving formulas on
two-partner games in ${\cal T}$ with more than 160 nested
alternated quantifiers.

\end{abstract}

\section{Introduction}
The algebra of (possibly) infinite trees plays a fundamental role
in computer science: it is a model for composed data known as
{\itshape record} in Pascal or {\itshape structure} in C. The
construction operation corresponds to the creation of a new
record, i.e. of a cell containing  elementary information
possibly followed by $n$ cells, each one pointing to a record.
Infinite trees correspond to a circuit of pointers.

As early as 1976, G. Huet gave an algorithm for unifying
infinite terms, that is solving equations in that
algebra~\cite{hue}. K.L. Clark   proposed a complete axiomatization  of the equality theory, also called  Clark equational theory CET, and gave intuitions about a complete axiomatization of the theory of finite trees \cite{clark}.  B. Courcelle has studied the properties of
infinite trees in the scope of recursive program
schemes~\cite{cou1,cou2}. A. Colmerauer  has described the
execution of Prolog II, III and IV programs in terms of solving
equations and disequations in that algebra
~\cite{col84,Colmerauer90,ben}. 

M. Maher has axiomatized all the cases by complete first-order
theories \cite{Maher}, i.e. he has introduced the theory ${\cal
T}$ of finite or infinite trees having an infinite set $\Ff$ of
functional symbols.  It is this theory which has  been the starting point 
of our works. After having studied  its  properties,  we have created
a new class of  complete theories that we call \emph{decomposable theories}
and have shown that a lot of theories used in fundamental computer
science are decomposable. We can cite for example: the theory of finite trees,  of infinite trees,  of finite or infinite trees \cite{moi3},  of
additive  rational or real numbers  with addition and subtraction,  of linear dense order  without endpoints,  of ordered additive rational or real  numbers  with addition, subtraction and a linear dense order relation without endpoints, of the combination of trees and ordered additive rational or real numbers \cite{moi2}, of the construction of trees on an ordered set \cite{moi1}, of the extension into trees of first-order theories  \cite{moi4} and many other combinations of fundamental theories.  

T. Dao whose works  focused on the theory of finite or infinite trees  has given a first version of a general algorithm solving first order formulas in finite or infinite trees  \cite{dao1} using  a basic simplification of quantified conjunctions of tree atomic formulas. Unfortunately, this  simplification holds only in the theory of finite or infinite trees  and can not be used in  theories  having completely different properties, such as the  theory of additive rational or real numbers. We have then generalized this result by introducing the term \emph{decomposable theories} \cite{moi1,moi2} and by showing that in each decomposable theory $T$, every  quantified conjunction of atomic  formulas can be decomposed  into  three embedded sequences of quantifications having
very particular properties, which can be expressed with the help
of three special quantifiers denoted by $\exists?$, $\exists!$,
$\exists^{\Psi(u)}_{\infty}$ and called \emph{at-most-one},
\emph{exactly-one}, \emph{infinite}. While the quantifiers $\exists?$, $\exists!$ are just
convenient notations already used in other works, the new quantifier $\exists^{\Psi(u)}_{\infty}$,
one of the essential keys of this class of theories, expresses a
property which is not expressible at the first-order level.

On the other hand, we wish to be  able to extract from the  definition of decomposable theory a general algorithm for solving first-order formulas  in any decomposable theory $T$. For that, we have  given  an efficient  algorithm for solving first-order formulas  in finite or infinite trees from which we have deduced  a general algorithm for solving first-order formulas  in any decomposable theory $T$  \cite{moi3}. Note that the first part\footnote{ The algorithm for solving first-order formulas  in finite or infinite trees.}  of  \cite{moi3}  was  a joint work with T. Dao in which we  improved the algorithm of \cite{dao1} and presented interesting benchmarks on finite or infinite trees with high performances. By solving a formula $\varphi$ (with or
without free variables) in a decomposable theory $T$, we mean to transform $\varphi$ into
a conjunction  $\phi$ of solved formulas, which is equivalent to
$\varphi$ in $T$, does not contain new free variables and such that: (1) either $\phi$ is the formula  $\vrai$, thus $\varphi$ is always true in $T$, (2) or
$\phi$ is the formula $\neg\vrai$, thus $\varphi$ is always false
in $T$, (3) or $\phi$ has at least one free variable and is easily
transformable into a Boolean combination of existentially
quantified conjunctions of atomic formulas. In particular, if
$\varphi$ has no free variables then $\phi$ is either the formula
$\vrai$ or $\neg\vrai$. 

Recently, we have also shown that an extension of the model of Prolog III and IV is possible by allowing the user to incorporate universal and existential quantifiers to  Prolog clauses and to solve any first-order formula, with or without free variables, in  a combination of trees and   first-order theories \cite{moi4}. For that, we have first  given an automatic way to combine any first-order theory $T$ with the theory of finite or infinite trees. Note that the two theories can have  non-joint signatures. Then, using the definition of decomposable theories, we have established simple conditions on $T$ and only on $T$ to get a decomposable combination and thus a complete combination. These extended theories have an interesting power of expressiveness and allow us to model complex problems with first-order formulas in  a combination of trees and other first-order theories. We can cite for example the works of Alain Colmerauer \cite{Colmerauer90} who has described the execution of Prolog III using a combination of trees and rational numbers  with addition, subtraction and linear dense order relation. A full proof of the decomposability of this hybrid  theory  can be found in detail  in \cite{moi2}.

The paper is organized in five sections followed by a conclusion.
This introduction is the first section. The second one introduces
the needed  elements of first-order logic and ends with a 
sufficient condition for the completeness of any first-order
theory. We have built this condition using a syntactic analysis of the 
 general structure of  first-order formulas.

In   section 3, we present the vectorial quantifiers $\exists?$, $\exists!$,
$\exists^{\Psi(u)}_{\infty}$ and show some of their properties. We also give  a formal  definition of  \emph{
decomposable theories} and show their completeness using the sufficient condition of completeness defined in section 2. If $T$ is decomposable, we show  that each formula is equivalent in $T$ to a Boolean combination of basic formulas and give a sufficient condition so that $T$  accepts full elimination of quantifiers.  We end this section with two examples of simple decomposable theories: a simple extension of the Clark equational theory CET \cite{clark} and the theory of  rational or real numbers  with addition and subtraction.

In  section 4, we present our algorithm of resolution in any
decomposable theory $T$, given in the  form of a set of five
rewriting rules. The conjunction $\phi$ of solved formulas obtained from an
initial formula $\varphi$ is equivalent to $\varphi$ in $T$ and
does not have new free variables. In particular, if $\varphi$ has
no free variables then $\phi$ is either the formula $\vrai$ or
$\neg\vrai$.  The correctness of our algorithm is another proof of completeness of the decomposable theories.  

Finally, we show in section 5 that the theory ${\cal T}$ of
finite or infinite trees is a decomposable theory and end with
examples and benchmarks done by an implementation of our algorithm
solving formulas on two-partner games in ${\cal T}$ with more
than 160 nested alternated quantifiers. We compare our results with those of \cite{moi3}, \cite{dao1} and \cite{dao2} where a dedicated algorithm for solving finite or infinite tree constraints has been given. We show that we have competitive results even if our algorithm is general and holds for any decomposable theory $T$.  This is a detailed full version with full proofs of our works on decomposable theories \cite{moi2,moi3}. The infinite quantifier, the properties of the vectorial quantifiers, the 
 class of the decomposable theories and the algorithm of resolution in any decomposable theory are our  contributions in all these works. The proof of decomposability of the theory of equality and the theory of additive rational or real numbers as well as the benchmarks on decomposable theories are  our main contributions in this paper.  

\section{Formal preliminaries\label{logique}}
\subsection{Expression \label{langage}}
We are given once and for all, an infinite countable set $\Vv$ of {\em
variables} and the set $L$ of \emph{logical} symbols:
\[=, \vrai, \faux,\neg ,\wedge, \vee, \rightarrow, \leftrightarrow, \forall,\exists, (, ).\]

We are also given once and for all, a \emph{signature} $S$, i.e. a set
of symbols partitioned into two subsets: the set of
\emph{function} symbols and the set of \emph{relation} symbols. To
each element $s$ of $S$ is linked a non-negative integer called
\emph{arity} of $s$. An $n$-ary symbol is a symbol with arity $n$.
A $0$-ary function symbol is called \emph{constant}.

As usual, an \emph{expression} is a word on $L\cup S\cup V$ which is
either a \emph{term}, i.e. of one of the two forms:
\begin{equation} \label{terme}
x,\;ft_1\ldots t_n,
\end{equation}
or a \emph{formula}, i.e. of  one of the eleven forms:
\begin{equation}\label{formules}\begin{array}{@{}c@{}}
s= t,\; rt_1\ldots t_n,\; \vrai,\; \faux,\;  \\
\neg \varphi,\;(\varphi \wedge \psi),\; (\varphi \vee
\psi),\;(\varphi \rightarrow \psi),\;(\varphi \leftrightarrow
\psi),\\(\forall x\,\varphi),\;(\exists x\,\varphi).
\end{array}
\end{equation}
In (\ref{terme}), $x$ is taken from $\Vv$, $f$ is an $n$-ary
function symbol taken from $S$ and the $t_i$'s are shorter terms.
In (\ref{formules}), $s,t$ and the $t_i$'s are terms, $r$ is an
$n$-ary relation symbol taken from $S$ and $\varphi$ and $\psi$
are shorter formulas. The set of the expressions forms   \emph{a first-order
language with equality}.

The formulas of the first line of (\ref{formules}) are known as \emph{atomic}, and
\emph{flat} if they are of one of the following forms:

\[\vrai,\; \faux,\; x_0=x_1, x_0=fx_1...x_n,\; rx_1...x_n,\]
where all the $x_i$'s are possibly non-distinct  variables taken from $\Vv$, $f$ is an $n$-ary function
symbol taken from $S$ and $r$ is an $n$-ary relation symbol taken
from $S$. An \emph{equation} is a formula of the form $s=t$ with $s$ and $t$ terms.

An occurrence of a variable $x$ in a formula is
\emph{bound} if it occurs in a sub-formula of the form $(\forall
x\, \varphi)$ or $(\exists x\, \varphi)$. It is \emph{free} in the
contrary case. The \emph{free variables of a formula} are those
which have at least one free occurrence in this formula. A
\emph{proposition} or a \emph{sentence} is a formula without free
variables.  If $\varphi$ is a formula, then we denote by $var(\varphi)$ the set of the free variables of $\varphi$.

The syntax of the formulas being constraining, we allowed
ourselves to use infix notations for the binary symbols and to add
and remove brackets when there are no ambiguities.  

We do not distinguish two
formulas which can be made equal using the following
transformations of  sub-formulas:
\[\begin{array}{@{}c@{}}\varphi\wedge\varphi\Longrightarrow\varphi,\;\;\varphi\wedge\psi\Longrightarrow\psi\wedge\varphi,\;\;(\varphi\wedge\psi)\wedge\phi\Longrightarrow\varphi\wedge(\psi\wedge\phi),\\\varphi\wedge\vrai\Longrightarrow\varphi,\;\;\varphi\vee\faux\Longrightarrow\varphi.\end{array}\]

If $I$ is the set $\{i_1,...,i_n\}$, we call \emph{conjunction} of
formulas and write $\bigwedge_{i\in I}\varphi_i$, each formula of
the form
$\varphi_{i_1}\wedge\varphi_{i_2}\wedge...\wedge\varphi_{i_n}\wedge\vrai$.
In particular, for $I=\emptyset$, the conjunction $\bigwedge_{i\in
I}\varphi_i$ is reduced to $\vrai$.   We denote by $FL$ the set of the conjunctions of flat formulas.  We denote by $AT$ the set of the conjunctions of atomic formulas.  A set $\Psi$ of formulas is \emph{closed under conjunction} if for each formula $\varphi\in \Psi$ and each formula $\phi\in \Psi$, the formula $\varphi\wedge\phi$  belongs to $\Psi$.  All theses considerations will be  useful for the algorithm of
resolution given in section 4.

\subsection{Model}
A \emph{model} is a couple $M = ({\mathcal M},\mathcal F)$, where:
\begin{itemize}
\item $\mathcal M$, the \emph{universe} or \emph{domain} of $M$,
is a {\bfseries nonempty set} disjoint from $S$, its elements are called
\emph{individuals} of $M$; \item $\mathcal F$ is a family of
operations and relations in the set $\mathcal M$, subscripted by
the elements of $S$ and such that: \begin{itemize}\item for every
$n$-ary function symbol $f$ taken from $S$, $f^M$ is an $n$-ary
operation in $\mathcal M$, i.e. an application from ${\mathcal
M}^n$ in ${\mathcal M}$. In particular, when $f$ is a constant,
$f^M$ belongs to ${\mathcal M}$; \item for every $n$-ary relation
symbol $r$ taken from $S$, $r^M$ is an $n$-ary relation in
${\mathcal M}$, i.e. a subset of ${\mathcal M}^n$.
\end{itemize}\end{itemize}

Let $M = ({\mathcal M},\mathcal F)$ be a model. An
\emph{$M$-expression} $\varphi$ is an expression built on the
signature $S\cup\mathcal{M}$ instead of $S$, {by considering the
elements of $\mathcal {M}$ as $0$-ary function symbols}. If for
each free variable $x$ of $\varphi$, we replace each free
occurrence of $x$ by a same element in $\mathcal{M}$, we get an
$M$-expression called \emph{instantiation} or \emph{valuation} of $\varphi$ by
individuals of $M$.

 If $\varphi$ is an $M$-formula, we say that
$\varphi$ \emph{is true in $M$} and we write
\begin{equation}\label{model}
M\models\varphi,
\end{equation}
if for any instantiation $\varphi^\prime$ of $\varphi$ by
individuals of $M$, the set $\mathcal M$ has the property
expressed by $\varphi\prime$, when we interpret the function and
relation symbols of $\varphi\prime$ by the corresponding functions
and relations of $M$ and when we give to the logical symbols their
usual meaning.
\begin{remarque} \label{modelecomplet} For every $M$-formula $\varphi$ without free variables, one and only one of the following properties holds: $M\models \varphi$, $M \models \neg \varphi$.
\end{remarque}

Let us finish this sub-section by a convenient notation. Let
$\bar{x}=x_1...x_n$ be a word on $\Vv$ and let $\bar{i}=i_1...i_n$
be a word on $\mathcal M$ or $\Vv$ of the same length as
$\bar{x}$. If $\varphi(\bar{x})$ and $\phi$ are two $M$-formulas, then we denote
by $\varphi(\bar{i})$, respectively $\phi_{\bar{x}\leftarrow\bar{i}}$ , the $M$-formula obtained by replacing in
$\varphi(\bar{x})$, respectively in $\phi$, each free occurrence of $x_j$ by $i_j$

\subsection{Theory}
A \emph{theory} is a (possibly infinite) set of propositions called \emph{axioms}. 
We say
that the model $M$ is a \emph{model of $T$}, if for each element
$\varphi$ of $T$, $M\models\varphi$. If $\varphi$ is a formula, we
write
\[ T\models\varphi,\]
if for each model $M$ of $T$, $M\models\varphi$. We say that the
formulas $\varphi$ and $\psi$ are \emph{equivalent in} $T$ if $T
\models \varphi\leftrightarrow \psi$. 

Let $T$ be a theory. A set $\Psi$ of formulas is called $T$-closed if: \begin{itemize}\item $\Psi\subseteq AT$, \item  $\Psi$ is closed under conjunction, \item every flat formula $\varphi$ is equivalent in $T$ to a formula which belongs to $\Psi$ and does
not contain other free variables than those of $\varphi$.
 \end{itemize}
The sets $AT$ and $FL$ are $T$-closed in any theory $T$. This notion of $T$-closed set is useful when we need to transform formulas of $FL$ into formulas which belong to  $\Psi$.  The transformation of normalized formulas into working formulas defined at Section \ref{hihiho1} illustrates this notion. 

A theory $T$ is \emph{complete} if for every proposition
$\varphi$, one and only one of the following properties holds:
$T\models \varphi$, $T \models \neg \varphi$.

Let us now present a  sufficient condition for the
completeness of any first-order theory. We will use the
abbreviation wnfv for \emph{``without new free variables }". A
formula $\varphi$ is equivalent to a wnfv formula $\psi$ in $T$
means that $T\models\varphi\leftrightarrow\psi$ and $\psi$ does
not contain other free variables than those of $\varphi$.

\begin{propriete}\label{cond}
A theory $T$ is complete if there exists a set of formulas, called
\emph{basic formulas}, such that:
\begin{enumerate}
\item every flat  formula is equivalent in $T$ to a wnfv
Boolean combination of basic formulas,
 \item every basic formula without free variables is equivalent in $T$, either to
 $\vrai$ or to $\faux$,
 \item every formula of the form
\begin{equation}
\textstyle{\exists x\,((\bigwedge_{i\in
I}\varphi_i)\wedge(\bigwedge_{i\in I'}\neg\varphi_i)),}
\end{equation}
where the $\varphi_i$'s are basic formulas, is equivalent in $T$
to a wnfv Boolean combination of basic formulas.
\end{enumerate}
\end{propriete}

\begin{proof} Let $\Phi$ be the set of all the formulas which are
equivalent in $T$ to a wnfv Boolean combination of basic formulas.

Let us show first that every formula $\psi$ belongs to $\Phi$. Let
us make a proof by induction on the syntactic structure of $\psi$. Without losing generalities we can restrict ourselves to the
cases where $\psi$ contains only flat formulas 	and the
following logical symbols\footnote{Because each atomic formula is equivalent in the empty theory to a wnfv quantified conjunction of flat formulas and each formula is equivalent in the empty theory to a wnfv formula which contains only the logical symbols: $\exists$, $\wedge$, $\neg$.}: $\neg$, $\wedge$, $\exists$. If $\psi$
is a flat  formula, then $\psi\in\Phi$ according to the
first condition of the property. If $\psi$ is of the form
$\neg\varphi_1$ or $\varphi_1\wedge\varphi_2$, with
$\varphi_1,\varphi_2\in\Phi$, then $\psi\in\Phi$ according to the
 definition of $\Phi$. If $\psi$ is of the form $\exists
x\,\varphi$, with $\varphi\in\Phi$, then according to the
definition of $\Phi$, the formula $\varphi$ is equivalent to a
wnfv formula $\varphi'$, which is a Boolean combination of basic
formulas $\varphi_{ij}$. Without losing generalities we can
suppose that $\varphi'$ is of the form
\begin{equation}
\textstyle{\varphi'=\bigvee_{i\in I}((\bigwedge_{j\in
J}\varphi_{ij})\wedge(\bigwedge_{j\in J'}\neg\varphi_{ij})).}
\end{equation}
By distributing the existential quantifier, the formula $\exists
x\,\varphi'$ is equivalent in $T$ to
\begin{equation}
\textstyle{\bigvee_{i\in I}(\exists x\,((\bigwedge_{j\in
J}\varphi_{ij})\wedge(\bigwedge_{j\in J'}\neg\varphi_{ij}))),}
\end{equation}
which, according to the third condition of the property, belongs
to $\Phi$. Thus the formula $\exists x\,\varphi$, i.e. $\psi$,
belongs to $\Phi$.

Let now $\psi$ be a proposition. According to what we have just
shown $\psi\in\Phi$. Thus, the formula $\psi$ is equivalent in $T$
to a Boolean combination of basic formulas without free variables.
According to the second condition of the property, one and only
one of the following properties holds: $T\models\psi$,
$T\models\neg\psi$. Thus $T$ is a complete theory.\end{proof}
This sufficient condition is interesting in the sense that it reasons on the syntactic structure of first-order formulas. Informally, the basic formulas are generally formulas of the form $\exists\bar{x}\alpha$ with $\alpha\in AT$. We will use this sufficient condition  in Section \ref{decc} to show the completeness of the decomposable theories. 

\begin{corollaire}\label{bono}
If $T$ satisfies the three conditions of Property \ref{cond} then every formula is equivalent in $T$ to a wnfv Boolean combination of basic formulas.
\end{corollaire}
This corollary  is a consequence of the proof of Property \ref{cond} in which  we have shown that if  $\Phi$ is the set of all the formulas which are
equivalent in $T$ to a wnfv Boolean combination of basic formulas then every formula $\psi$ belongs to $\Phi$. 

\section{Decomposable theory}

\subsection{Vectorial quantifiers \label{besoins}}

Let $M$ be a model and let $T$ be a theory. Let $\bar{x}=x_1\ldots
x_n$ and $\bar{y}=y_1\ldots y_n$ be two words on $\Vv$ of the same
length. Let  $\phi$, $\varphi$ and $\varphi(\bar{x})$ be
$M$-formulas. We write
\[\begin{tabular}{@{}l@{\,\,}l@{\,\,}l@{}}
$\exists\bar{x}\,\varphi$&for&$\exists x_1...\exists x_n\,\varphi$,\\
$\forall\bar{x}\,\varphi$&for&$\forall x_1...\forall x_n\,\varphi$,\\
$\exists?\bar{x}\,\varphi(\bar{x})$&for&$\forall\bar{x}\forall\bar{y}\,\varphi(\bar{x})\wedge\varphi(\bar{y})\rightarrow \bigwedge_{i\in\{1,...,n\}}x_i=y_i,$\\
$\exists!\bar{x}\,\varphi$&for&$(\exists\bar{x}\,\varphi)\wedge(\exists?\bar{x}\,\varphi).$
\end{tabular}\]
The word $\bar{x}$, which can be the empty word $\varepsilon$, is
called \emph{vector of variables}. Note that the formulas
$\exists?\varepsilon\varphi$ and $\exists!\varepsilon\varphi$ are
respectively equivalent to $\vrai$ and to $\varphi$ in any model
$M$. 

\begin{notation} Let $Q$ be a  quantifier taken from $\{\forall,\exists, \exists!, \exists?\}$. Let $\bar{x}$ be vector of  variables taken from $V$.  We write:   \[Q\bar{x}\,\varphi\wedge\phi\;\; for\;\; Q\bar{x}\,(\varphi\wedge\phi).\] 
\end{notation}
\begin{example}
Let $I=\{1,...,n\}$ be a finite  set. Let $\varphi$ and $\phi_i$ with $i\in I$ be formulas. Let $\bar{x}$ and $\bar{y}_i$ with $i\in I$ be vectors of variables. We write:
\[\begin{tabular}{@{}l@{\,\,}l@{\,\,}l@{}}
$\exists\bar{x}\,\varphi\wedge\neg\phi_1$&for&$\exists\bar{x}\,(\varphi\wedge\neg\phi_1)$,\\
$\forall\bar{x}\,\varphi\wedge\phi_1$&for&$\forall\bar{x}\,(\varphi\wedge\phi_1)$,\\
$\exists!\bar{x}\,\varphi\wedge\bigwedge_{i\in I}(\exists\bar{y}_i\phi_i)$&for&$\exists!\bar{x}\,(\varphi\wedge(\exists\bar{y}_1\phi_1)\wedge...\wedge(\exists\bar{y}_n\phi_n)\wedge\vrai),$\\
$\exists?\bar{x}\,\varphi\wedge\bigwedge_{i\in I}\neg(\exists\bar{y}_i\phi_i)$&for&$\exists?\bar{x}\,(\varphi\wedge(\neg(\exists\bar{y}_1\phi_1))\wedge...\wedge(\neg(\exists\bar{y}_n\phi_n))\wedge\vrai).$
\end{tabular}\]
\end{example}

\begin{propriete}\label{attention}
If $T\models\, \exists?\bar{x}\,\varphi$ then \begin{equation}\label{eq1.2} T\models\,
(\exists\bar{x}\,\varphi \wedge
\neg\phi)\leftrightarrow((\exists\bar{x}\varphi)\wedge\neg(\exists\bar{x}\,\varphi\wedge\phi)).\end{equation}
\end{propriete}

\begin{proof}
Let  $M$ be a model of $T$ and let
$\exists\bar{x}\,\varphi'\wedge\neg\phi'$ be an instantiation of
$\exists\bar{x}\,\varphi\wedge\neg\phi$ by individuals of $M$. Let
us denote by $\varphi'_1$ the $M$-formula
$(\exists\bar{x}\,\varphi'\wedge\neg\phi')$ and by $\varphi'_2$
the $M$-formula
$(\exists\bar{x}\,\varphi')\wedge\neg(\exists\bar{x}\varphi'\wedge\phi')$.
To show the equivalence (\ref{eq1.2}), it is enough to show that
\begin{equation}\label{eq1.1}
M\models\varphi'_1\leftrightarrow\varphi'_2. \end{equation} If
$M\models\neg(\exists\bar{x}\,\varphi')$ then
$M\models\neg\varphi'_1$ and $M\models\neg\varphi'_2$, thus
the equivalence (\ref{eq1.1}) holds.\\
If $M\models\exists\bar{x}\,\varphi'$. Since $T\models\exists
?\bar{x}\,\varphi'$, there exists a unique vector $\bar{i}$ of
individuals of $M$ such that
$M\models\varphi'_{\bar{x}\leftarrow\bar{i}}$. Two cases arise:

If $M\models\neg(\phi'_{\bar{x}\leftarrow\bar{i}})$, then
$M\models(\varphi'\wedge\neg \phi')_{\bar{x}\leftarrow\bar{i}}$,
thus $M\models\varphi'_1$. Since $\bar{i}$ is unique and since
$M\models\neg(\phi'_{\bar{x}\leftarrow\bar{i}})$, there exists no
vector $\bar{u}$ of individuals of $M$ such that
$M\models(\varphi'\wedge\phi')_{\bar{x}\leftarrow\bar{u}}$.
Consequently, $M\models\neg(\exists\bar{x}\,\varphi'\wedge\phi')$
and thus $M\models\varphi'_2$. We have $M\models\varphi'_1$ and
$M\models\varphi'_2$, thus the equivalence (\ref{eq1.1}) holds.

If $M\models\phi'_{\bar{x}\leftarrow\bar{i}}$, then
$M\models(\varphi'\wedge\phi')_{\bar{x}\leftarrow\bar{i}}$ and
thus $M\models\neg\varphi'_2$. Since $\bar{i}$ is unique
 and since $M\models\phi'_{\bar{x}\leftarrow\bar{i}}$, there exists no vector $\bar{u}$ of
 individuals of $M$ such that $M\models(\varphi'\wedge\neg\phi')_{\bar{x}\leftarrow\bar{u}}$. Consequently,
 $M\models\neg(\exists\bar{x}\,\varphi'\wedge\neg\phi')$
and thus $M\models\neg\varphi'_1$. We have
$M\models\neg\varphi'_1$ and $M\models\neg\varphi'_2$, thus  the
equivalence (\ref{eq1.1}) holds.\end{proof} 

\begin{corollaire}\label{r2}
If $T\models$\, $\exists?\bar{x}\,\varphi$ then  \[T\models\,
(\exists\bar{x}\,\varphi \wedge\bigwedge_{i\in I}
\neg\phi_i)\leftrightarrow((\exists\bar{x}\varphi)\wedge\bigwedge_{i\in I}\neg(\exists\bar{x}\,\varphi\wedge\phi_i)).\]
\end{corollaire}

\begin{proof}
Let $\psi$ be the formula $\neg(\bigwedge_{i\in I}\neg\phi_i)$. The formula 
$\exists\bar{x}\,\varphi \wedge\bigwedge_{i\in I}
\neg\phi_i,
$
is equivalent in $T$ to
$
\exists\bar{x}\,\varphi \wedge\neg\psi.
$
Since $T\models\exists?\bar{x}\,\varphi$, then according to Property \ref{attention} the preceding formula is equivalent in $T$ to
 $
(\exists\bar{x}\,\varphi) \wedge\neg(\exists\bar{x}\,\varphi\wedge\psi),
$
which is equivalent in $T$ to
 $
(\exists\bar{x}\,\varphi) \wedge\neg(\exists\bar{x}\,\varphi\wedge\neg(\bigwedge_{i\in I}\neg\phi_i)),
$
thus to
 $
(\exists\bar{x}\,\varphi) \wedge\neg(\exists\bar{x}\,\varphi\wedge(\bigvee_{i\in I}\phi_i)),
$
which is equivalent in $T$ to
 $
(\exists\bar{x}\,\varphi) \wedge\neg(\exists\bar{x}\,(\bigvee_{i\in I}(\varphi\wedge\phi_i))),
$
thus to
 $
(\exists\bar{x}\,\varphi) \wedge\neg(\bigvee_{i\in I}(\exists\bar{x}\,\varphi\wedge\phi_i)),
$
which is finally equivalent in $T$ to 
 $
(\exists\bar{x}\,\varphi) \wedge\bigwedge_{i\in I}\neg(\exists\bar{x}\,\varphi\wedge\phi_i).
$
 \end{proof}

\begin{propriete}\label{zebbb}
If $T\models\, \exists!\bar{x}\,\varphi$ then
\[T\models(\exists\bar{x}\,\varphi \wedge
\neg\phi)\leftrightarrow\neg(\exists\bar{x}\,\varphi\wedge\phi).\]
\end{propriete}

\begin{corollaire}\label{unique1}
If $T\models\exists!\bar{x}\,\varphi$ then
\[T\models(\exists\bar{x}\,\varphi \wedge\bigwedge_{i\in I}
\neg\phi_i)\leftrightarrow\bigwedge_{i\in I}\neg(\exists\bar{x}\,\varphi\wedge\phi_i).\]
\end{corollaire}

\subsection{The infinite quantifier }
Let $M$ be a model. Let $T$ be a theory. Let $\varphi(x)$ be an 
$M$-formula and let $\Psi(u)$ be a set of formulas having at most
$u$ as  free variable. Let us now present our infinite quantifier $\exists^{\Psi(u)}_{\infty}$. The main intuitions behind this quantifier come from an aim to get a full elimination of quantifiers in complex $M$-formulas of the form $\exists x\,\varphi(x)\wedge\bigwedge_{j\in
\{1,...,n\}}\neg\psi_j(x)$ using the fact that the domain of $M$ is infinite.  

\begin{definition1}\label{infini}
We write
\begin{equation}\label{inf}M\models\exists^{\Psi(u)}_{\infty}x\,\varphi(x),\end{equation}
if for every instantiation $\exists x\,\varphi'(x)$ of $\exists
x\,\varphi(x)$ by individuals of $M$ and for every finite subset
$\{\psi_1(u),..,\psi_n(u)\}$ of elements of $\Psi(u)$, the set of
the individuals $i$ of $M$ such that
$M\models\varphi'(i)\wedge\bigwedge_{j\in
\{1,...,n\}}\neg\psi_j(i)$ is infinite. $\\$ We write
$T\models\exists^{\Psi(u)}_{\infty}x\,\varphi(x),$ if for each
model $M$ of $T$ we have (\ref{inf}).
\end{definition1}
This infinite quantifier holds only for models whose set of individuals is infinite.  Note that if $\Psi(u)=\{\faux\}$ then (\ref{inf}) simply means that $M$ contains an infinite set of individuals $i$ such that $\varphi(i)$. 
Informally, the notation (\ref{inf})  states that there exists a full elimination of quantifiers in  formulas of the form 
$\exists x\,\varphi(x)\wedge\bigwedge_{j\in
\{1,...,n\}}\neg\psi_j(x)$ due to  an infinite set of  valuations of $x$ in $M$ which satisfy this formula. 

\begin{propriete}\label{linfini}
Let $J$ be a finite (possibly empty) set. Let $\varphi(x)$ and
$\varphi_j(x)$ with $j\in J$ be $M$-formulas. If
 $T\models\exists^{\Psi(u)}_{\infty}x\,\varphi(x)$ and if
for each $\varphi_j(x)$, at least one of the following properties
holds:
\begin{itemize}
\item $T\models\exists?x\,\varphi_j(x)$,
 \item
there exists $\psi_j(u)\in\Psi(u)$ such that $T\models\forall
x\,\varphi_j(x)\rightarrow\psi_j(x),$
\end{itemize}
then
\[
\textstyle{T\models\exists x\,\varphi(x)\wedge\bigwedge_{j\in
J}\neg\varphi_j(x)}
\]
\end{propriete}

\begin{proof}
Let $M$ be a model of $T$ and let $\exists
x\,\varphi'(x)\wedge\bigwedge_{j\in J}\neg\varphi'_j(x)$ be an
instantiation of $\exists x\,\varphi(x)\wedge\bigwedge_{j\in
J}\neg\varphi_j(x)$ by individuals of $M$. Suppose that the
conditions of Property \ref{linfini} hold and let us show that
\begin{equation}\label{equiv15}
\textstyle{M\models\exists x\,\varphi'(x)\wedge\bigwedge_{j\in
J}\neg\varphi'_j(x)}.
\end{equation}
Let $J'$ be the set of the $j\in J$ such that
$M\models\exists?x\,\varphi'_j(x)$ and let $m$ be the cardinality
of $J'$. Since for all $j\in J'$,
$M\models\exists?x\,\varphi'_j(x)$, then for every set $\mathcal M'$
of individuals of $M$ such that $Cardinality(\mathcal M')>m$,
there exists $i\in \mathcal M'$ such that
\begin{equation}\label{cy1} M\models\,\bigwedge_{j\in
J'}\neg\varphi'_j(i).\end{equation} On the other hand, since
$T\models\exists^{\Psi(u)}_{\infty}x\,\varphi(x)$ and according to
Definition \ref{infini} we know that for every finite subset
$\{\psi_1(u),...,\psi_n(u)\}$ of $\Psi(u)$, the set of the
individuals $i$ of $M$ such that
$M\models\varphi'(i)\wedge\bigwedge_{k=1}^{n}\neg\psi_k(i)$ is
infinite. Since for all $j\in J-J'$ we have  $M\models\forall
x\,\varphi_j(x)\rightarrow\psi_j(x)$, thus $M\models\forall
x\,(\neg\psi_j(x))\rightarrow(\neg\varphi_j(x))$, then there
exists an infinite set $\xi$ of individuals $i$ of $M$ such that
$M\models\varphi'(i)\wedge\bigwedge_{j\in J-J'}\neg\varphi'_j(i)$.
Since $\xi$ is infinite then $Cardinality(\xi)>m$, and thus
according to (\ref{cy1}) there exists at least an individual $i\in
\xi$ such that $M\models\varphi'(i)\wedge(\bigwedge_{j\in
J-J'}\neg\varphi'_j(i))\wedge(\bigwedge_{k\in
J'}\neg\varphi'_k(i))$. Thus, we have $M\models\exists
x\,\varphi'(x)\wedge\bigwedge_{j\in J}\neg\varphi'_j(x).$
\end{proof}

\begin{propriete}\label{zebb}
If $T\models\exists^{\Psi(u)}_{\infty}x\,\varphi(x)$ then
$T\models\exists^{\Psi(u)}_{\infty}x\,\vrai$.
\end{propriete}
\begin{proof}
Let $M$ be a model of $T$. If  $T\models\exists^{\Psi(u)}_{\infty}x\,\varphi(x)$ then $M\models\exists^{\Psi(u)}_{\infty}x\,\varphi(x)$. According to Definition \ref{infini} there exists an infinite set of individuals $i$ such that $M\models\varphi(i)\wedge\bigwedge_{j\in
J}\neg\varphi_j(i)$ with $\varphi_j(u)\in\Psi(u)$ for all $j\in J$. Thus there exists an infinite set of individuals $i$ such that  $M\models\vrai\wedge\bigwedge_{j\in
J}\neg\varphi_j(i)$, i.e. $M\models\exists^{\Psi(u)}_{\infty}x\,\vrai$ and thus $T\models\exists^{\Psi(u)}_{\infty}x\,\vrai$.
\end{proof}

\subsection{Decomposable theory}\label{decc}

We present in this section a formal definition of the \emph{decomposable theories}. Informally, this definition simply states that in every decomposable theory $T$  each formula of the form $\exists\bar{x}\alpha$ with $\alpha$ a $T$-closed set is equivalent in $T$ to a decomposed formula of the form  $\exists \bar{x}'\,\alpha'\wedge(
\exists\bar{x}''\,\alpha''\wedge(\exists\bar{x}'''\,\alpha'''))$, where the formulas $\exists \bar{x}'\,\alpha'$, $\exists \bar{x}''\,\alpha''$ and $\exists \bar{x}'''\,\alpha'''$ have elegant properties which can be expressed using  vectorial quantifiers. 

\begin{definition1} \label{decomp}
A theory $T$ is called \emph{decomposable} if there exists a set
$\Psi(u)$ of formulas having at most $u$ as free variable, a $T$-closed set $A$  and
three sets $A'$, $A''$ and $A'''$ of formulas of the form
$\exists\bar{x}\alpha$ with $\alpha\in A$ such that:
\begin{enumerate}

 \item
Every formula of the form $\exists\bar{x}\,\alpha\wedge\psi$, with
$\alpha\in A$ and $\psi$ any formula, is equivalent in $T$ to a
wnfv decomposed formula of the form
\[
\exists \bar{x}'\,\alpha'\wedge(
\exists\bar{x}''\,\alpha''\wedge(\exists\bar{x}'''\,\alpha'''\wedge
\psi)),
\]
  with $\exists\bar{x}'\,\alpha'\in A'$, $\exists\bar{x}''\,\alpha''\in A''$ and
$\exists\bar{x}'''\,\alpha'''\in A'''$.

 \item If $\exists\bar{x}'\alpha'\in
A'$ then $T\models\exists?\bar{x}'\,\alpha'$ and for each free
variable $y$ in $\exists\bar{x}'\alpha'$, at least one of the
following properties holds:
\begin{itemize}
\item $T\models\exists?y\bar{x}'\,\alpha'$, \item there exists
$\psi(u)\in\Psi(u)$ such that $T\models\forall
y\,(\exists\bar{x}'\,\alpha')\rightarrow\psi(y)$.
\end{itemize}

\item If $\exists\bar{x}''\alpha''\in A''$ then for each $x''_i$
of $\bar{x}''$ we have
$T\models\exists^{\Psi(u)}_{\infty}x''_i\,\alpha''$.

\item If $\exists\bar{x}'''\alpha'''\in A'''$ then
$T\models\exists!\bar{x}'''\,\alpha'''$.

\item If the formula $\exists\bar{x}'\alpha'$ belongs to $A'$ and
 has no free variables then this formula is either the formula $\exists\varepsilon\vrai$
 or $\exists\varepsilon\faux$.

\end{enumerate}
\end{definition1}

Since $A$ is $T$-closed, then $A$ is a subset of $AT$.  While the formulas of  $A''$ and $A'''$ accept  full elimination of quantifiers according to the properties of the quantifiers $\exists!$ and $\exists^{\Psi(u)}_{\infty}$,  the formulas of $A'$ can possibly  not accept elimination of quantifiers. This is due to the second point of Definition \ref{decomp}  which states that $T\models\exists?\bar{x}'\alpha'$. The computation of the  sets $A$, $A'$, $A''$, $A'''$ and $\Psi(u)$ for a theory $T$ depends on the axiomatization of $T$.  Generally, it is enough to know how to solve a formula of the form $\exists\bar{x}\alpha$ with $\alpha\in FL$ to get a first intuition on the sets $A'$, $A''$, $A'''$ and $\Psi(u)$. Informally, the sets $A'$, $A''$ and $A'''$ can be called according to their linked vectorial quantifier, i.e. $A'$ is the \emph{at most one solution set} and contains formulas which accept at most one solution in $T$ and  possibly  not accept full elimination of quantifiers, the set $A''$ is the \emph{infinite instantiation set} and contains formulas that accept an infinite set of solutions in $T$. The set $A'''$ is the \emph{unique solution set} and contains  formulas which have one and only solution in $T$. The set $\Psi(u)$ contains generally  simple formulas of the form $\exists\bar{x}\alpha$ with at most one free variable and $\alpha\in A$. It can also be reduced for example to the set $\{faux\}$. Note  that the sets $A'$ and $A'''$ are generally not empty since for every model $M$ of any theory $T$ we have $M\models\exists?\varepsilon\,x=y$ and $M\models\exists! x\,x=y$. 

\begin{propriete}\label{dec2}
Let $T$ be a decomposable theory. Every formula of the form
$\exists\bar{x}\,\alpha,$ with $\alpha\in A$, is equivalent in $T$
to a wnfv formula of the form $\exists\bar{x}'\,\alpha'$ with
$\exists\bar{x}'\alpha'\in A'$.
\end{propriete}

\begin{proof}
Let $\exists\bar{x}\,\alpha$ be a formula with $\alpha\in A$.
According to Definition \ref{decomp} this formula is equivalent in
$T$ to a wnfv formula of the form
\[
\exists \bar{x}'\,\alpha'\wedge(
\exists\bar{x}''\,\alpha''\wedge(\exists\bar{x}'''\,\alpha''')),
\]
  with $\exists\bar{x}'\,\alpha'\in A'$, $\exists\bar{x}''\,\alpha''\in A''$ and
$\exists\bar{x}'''\,\alpha'''\in A'''$. Since
$\exists\bar{x}'''\,\alpha'''\in A'''$ then according to Definition
\ref{decomp} we have $T\models\exists!\bar{x}'''\alpha'''$ and thus using
Property  \ref{zebbb} (with $\phi=\faux$) the preceding  formula
is equivalent in $T$ to
\[
\exists \bar{x}'\,\alpha'\wedge( \exists\bar{x}''\,\alpha''),
\]
which is equivalent in $T$ to
\[
\exists \bar{x}'\,\alpha'\wedge( \exists
x''_1...x''_{n-1}\,(\exists x''_n\,\alpha'')).
\]
 Since
$\exists\bar{x}''\,\alpha''\in A''$ then according to Definition
\ref{decomp}  we have
$T\models\exists^{\Psi(u)}_{\infty}x''_n\,\alpha''$ and thus
$T\models\exists\,x''_n\,\alpha''$. The preceding  formula is
equivalent in $T$ to
\[
\exists \bar{x}'\,\alpha'\wedge( \exists x''_1...x''_{n-1}
\,\vrai),
\]
which  is finally 
equivalent in $T$ to \[ \exists \bar{x}'\,\alpha'.
\]
\end{proof}

Using Property \ref{dec2} and the fifth point of Definition
\ref{decomp} we get
\begin{corollaire}\label{bien}
Let $T$ be a decomposable theory. Every formula, without free
variables, of the form $\exists\bar{x}\,\alpha,$ with $\alpha\in
A$, is equivalent in $T$ either to $\vrai$ or to
$\faux$.\end{corollaire}

\begin{theoreme}\label{completude}
If $T$ is decomposable then $T$ is complete.
\end{theoreme}
\begin{proof}
Let $T$ be a decomposable theory which satisfies the five
conditions of Definition \ref{decomp}. Let us show that $T$ is
complete using  Property \ref{cond} and by taking  formulas
of the form $\exists\bar{x}\,\alpha$, with $\alpha\in A$, as basic
formulas. Note that according to Definition \ref{decomp}, the sets $A'$, $A''$ and $A'''$ contain formulas of the form
$\exists\bar{x}\alpha$ with $\alpha\in A$. 

Let us show that the first condition of  Property \ref{cond} holds, i.e. every flat formula is equivalent in $T$  to a wnfv Boolean combination of basic formulas. According to Definition \ref{decomp} the set 
$A$ is $T$-closed, i.e. (i)   every flat formula is equivalent in $T$ to a wnfv formula which belongs to $A$. Let $\alpha$ be a flat formula. According to (i) $\alpha$ is equivalent in $T$ to a wnfv formula $\beta$ which belongs to $A$. Since  $\beta$ is equivalent in $T$ to  $\exists\varepsilon\,\beta$ and $\beta\in A$ then $\alpha$ is equivalent in $T$ to a wnfv basic formula\footnote{Of course a basic formula is a particular case of a Boolean combination of basic formulas.}. Thus, the first condition of  Property \ref{cond} holds.

Let us show that the second condition of  Property \ref{cond} holds, i.e. every basic formula without free variables is either equivalent to $\vrai$ or to $\faux$ in $T$. Let $\exists\bar{x}\,\alpha$ with $\alpha\in A$ be a basic formula without free variables. According to Corollary \ref{bien} either $T\models\exists\bar{x}\alpha$ or $T\models\neg(\exists\bar{x}\,\alpha)$. Thus, the second condition of Property
\ref{cond} holds.

Let us show now that the third condition of  Property \ref{cond} holds, i.e. every formula of the form
\begin{equation}\label{add1}
\textstyle{\exists x\,(\bigwedge_{i\in
I}(\exists\bar{x}_i\,\alpha_i))\wedge(\bigwedge_{j\in
J}\neg(\exists\bar{y}_j\,\beta_j)),}
\end{equation}
with  $\alpha_i\in A$ for all $i\in I$ and $\beta_j\in A$ for all $j\in J$,  is equivalent in $T$ to a wnfv Boolean combination of
basic formulas, i.e. to a wnfv Boolean combination of formulas of
the form $\exists\bar{x}\alpha$ with $\alpha\in A$. By lifting all the
quantifications $\exists\bar{x}_i$ after having possibly renamed
the variables\footnote{We must rename the variables of $\bar{x}_i$ only if they have free occurrences in a formula $\alpha_k$ of (\ref{add1}) with $k\in I$ and $i\neq k$.} which appear in each $\bar{x}_i$, the formula (\ref{add1}) is equivalent in $T$ to a wnfv formula of the form
\[
\textstyle{\exists\bar{x}\,(\bigwedge_{i\in I}\alpha_i)\wedge\bigwedge_{j\in
J}\neg(\exists\bar{y}_j\,\beta_j),}
\]
with $\alpha_i\in A$ for all $i\in I$ and $\beta_j\in A$  for all $j\in J$. According to Definition \ref{decomp} the set $A$ is $T$-closed and thus closed under  
conjunction. The preceding formula is equivalent in $T$ to a wnfv formula of the form 
\[
\textstyle{\exists\bar{x}\,\alpha\wedge\bigwedge_{j\in
J}\neg(\exists\bar{y}_j\,\beta_j),}
\]
with $\alpha\in A$ and  $\beta_j\in A$  for all $j\in J$. According to the first point of Definition \ref{decomp}
the preceding formula is equivalent in $T$ to a wnfv formula of the form 
\[
\textstyle{\exists\bar{x}'\,\alpha'\wedge(\exists\bar{x}''\,\alpha''\wedge(\exists\bar{x}'''\,\alpha'''\wedge
\bigwedge_{j\in J}\neg(\exists\bar{y}_{j}\,\beta_{j}))),}
\]
with $\exists\bar{x}'\,\alpha'\in A'$, $\exists \bar{x}''\,\alpha''\in A''$, $\exists \bar{x}'''\,\alpha'''\in A'''$ and
$\beta_{j}\in A$ for all $j\in J$. Since $\exists \bar{x}'''\,\alpha'''\in A'''$ then according to the fourth point of Definition \ref{decomp} $T\models\exists !\bar{x}'''\,\alpha''$. Thus, using Corollary \ref{unique1}  the preceding formula is equivalent in $T$ to

\[
\textstyle{\exists\bar{x}'\,\alpha'\wedge(\exists\bar{x}''\,\alpha''\wedge
\bigwedge_{j\in J}\neg(\exists\bar{x}'''\,\alpha'''\wedge(\exists\bar{y}_{j}\,\beta_{j}))).}
\]
By lifting all the quantifies $\exists\bar{y}_{j}$ after having possibly renamed the variables which appear in each $\bar{y}_j$, the preceding formula is equivalent in $T$ to

\[
\textstyle{\exists\bar{x}'\,\alpha'\wedge(\exists\bar{x}''\,\alpha''\wedge
\bigwedge_{j\in J}\neg(\exists\bar{x}'''\exists\bar{y}_{j}\,\alpha'''\wedge\beta_{j})).}
\]
According to Definition \ref{decomp} the sets $A'$, $A''$ and $A'''$ contain formulas of the form $\exists\bar{x}\alpha$ with $\alpha\in A$, thus $\alpha'''\in A$. Since $\beta_j\in A$ for all $j\in J$ and  since $A$ is $T$-closed (i.e. closed under conjunction...) then for all $j\in J$ the formula $\alpha'''\wedge\beta_{j}$ belongs to $A$. Thus, the preceding formula is equivalent in $T$ to a wnfv formula of the form
\[
\textstyle{\exists\bar{x}'\,\alpha'\wedge(\exists\bar{x}''\,\alpha''\wedge
\bigwedge_{j\in J}\neg(\exists\bar{y}_{j}\,\beta_{j})),}
\]
 with  $\exists\bar{x}'\,\alpha'\in A'$, $\exists \bar{x}''\,\alpha''\in A''$,  and
$\beta_{j}\in A$ for all $j\in J$. According to Property \ref{dec2} the preceding formula is equivalent in $T$ to a wnfv formula of the form 
\[
\textstyle{\exists\bar{x}'\,\alpha'\wedge(\exists\bar{x}''\,\alpha''\wedge
\bigwedge_{j\in J}\neg(\exists\bar{y}'_{j}\,\beta'_{j})),}
\]
with  $\exists\bar{x}'\,\alpha'\in A'$, $\exists \bar{x}''\,\alpha''\in A''$,  and
$\exists\bar{y}'_j\,\beta'_{j}\in A'$ for all $j\in J$.
 Let us
denote by $J_{1}$, the set of the $j \in J$ such that $x''_n$ does
not have free occurrences in the formula
$\exists\bar{y}'_j\beta'_j$. Thus, the preceding formula is
equivalent in $T$ to
\begin{equation}\label{kh01}
\exists\bar{x}'\,\alpha'\wedge(\exists x''_1...\exists
x''_{n-1}\left[\begin{array}{@{}l@{}}(\bigwedge_{j\in
J_1}\neg(\exists\bar{y}'_j\,\beta'_j))\wedge\\(\exists
x''_n\,\alpha''\wedge\bigwedge_{j\in
J-J_1}\neg(\exists\bar{y}'_j\,\beta'_j))\end{array}\right]).
\end{equation}
Since 
$\exists\bar{x}''\,\alpha''\in A''$ and
$\exists\bar{y}'_j\,\beta'_j\in A'$ for all $j\in J$, then according to Property
\ref{linfini} and the points  2 and 3 of Definition
\ref{decomp}, the formula (\ref{kh01}) is equivalent in $T$ to
\[
\textstyle{\exists\bar{x}'\,\alpha'\wedge(\exists
x''_1...\exists x''_{n-1}\,(\vrai\wedge\bigwedge_{j\in
J_1}\neg(\exists\bar{y}'_j\,\beta'_j))).}
\]
By repeating the three preceding steps $(n-1)$ times, by denoting
by $J_{k}$ the set of the $j\in J_{k-1}$ such that $x''_{(n-k+1)}$
does not have free occurrences in $\exists\bar{y}'_j\,\beta'_j$, and
by using $(n-1)$ times Property \ref{zebb}, the preceding formula
is equivalent in $T$ to
\[
\textstyle{\exists\bar{x}'\,\alpha'\wedge\bigwedge_{j\in
J_n}\neg(\exists\bar{y}'_j\,\beta'_j).}
\]
Since $\exists\bar{x}'\,\alpha'\in A'$ then according to the second point of Definition \ref{decomp} we have $T\models\exists?\bar{x}'\,\alpha'$. Thus, using Corollary \ref{r2} the preceding formula is equivalent in $T$ to
\[
\textstyle{(\exists\bar{x}'\,\alpha')\wedge\bigwedge_{j\in
J_n}\neg(\exists\bar{x}'\,\alpha'\wedge(\exists\bar{y}'_j\,\beta'_j)).}
\]
By lifting all the quantifies $\exists\bar{y}_{j}$ after having possibly renamed the variables which appear in each $\bar{y}_j$, the preceding formula is equivalent in $T$ to 
\[
\textstyle{(\exists\bar{x}'\,\alpha')\wedge\bigwedge_{j\in
J_n}\neg(\exists\bar{x}'\exists\bar{y}'_j\,\alpha'\wedge\beta'_j).}
\]
According to Definition \ref{decomp} the sets $A'$, $A''$ and $A'''$ contain formulas of the form $\exists\bar{x}\alpha$ with $\alpha\in A$. Thus, since $\exists\bar{x}'\,\alpha'\in A'$ and $\exists\bar{y}'_j\,\beta'_j\in A'$ for all $j\in J_n$, then $\alpha'\in A$ and $\beta_j\in A$ for all $j\in J_n$. Since the set $A$ is $T$-closed, it is closed under conjunction, then for all $j\in J_n$  the formula $\alpha'\wedge\beta'_j$ belongs to $A$ and thus, the preceding formula is equivalent in $T$ a wnfv formula of the form
\[
\textstyle{(\exists\bar{x}\,\alpha)\wedge\bigwedge_{j\in
J_n}\neg(\exists\bar{y}_j\beta_j),}
\]
with $\alpha\in A$ and  $\beta_j\in A$ for all $j\in J_n$. This formula is a Boolean combination of formulas of the form $\exists\bar{x}\alpha$ with $\alpha\in A$, i.e. a Boolean combination of basic formulas.  Thus, the third condition of Property
\ref{cond} holds.

Since $T$ satisfies the three conditions of Property \ref{cond}, then $T$ is a complete theory. 
\end{proof}

According to Theorem \ref{completude} and Corollary \ref{bono}, we have the following corollary:
\begin{corollaire}\label{elimq}
If $T$ is decomposable and if for all $\exists\bar{x}'\alpha'\in A'$ we have $\bar{x}'=\varepsilon$, then $T$ accepts full elimination of quantifiers.
\end{corollaire}
\begin{proof}
Let $T$ be a decomposable theory such that for all $\exists\bar{x}'\alpha'\in A'$ we have $\bar{x}'=\varepsilon$.  Let $\varphi$ be a formula which can possibly contain free variables. In the proof of Theorem \ref{completude} we have shown that $T$ satisfies the three conditions of Property \ref{cond} using  formulas of the forms $\exists\bar{x}\alpha$ with $\alpha\in A$ as basic formulas. Thus, according to Corollary \ref{bono}, the formula $\varphi$ is equivalent in $T$ to a wnfv Boolean combination of basic formulas, i.e. Boolean combination of formulas of the form $\exists\bar{x}\alpha$ with $\alpha\in A$. According to Property \ref{dec2} each of these basic formulas is equivalent in $T$ to a wnfv formula of the form $\exists\bar{x}'\alpha'$ which belongs to $A'$.  Since for all $\exists\bar{x}'\alpha'\in A'$ we have $\bar{x}'=\varepsilon$ and  since $\alpha'\in A$ (according to the structure of the set $A'$ defined in Definition \ref{decomp}) then the formula $\varphi$ is equivalent in $T$ to a boolean combination of elements of $A$. Since $T$ is decomposable then $A$ is a  $T$-closed set and thus $A\subseteq AT$. Then, the formula $\varphi$ is  equivalent in $T$ to a boolean combination $\phi$ of conjunctions of atomic formulas. According to the syntax of the atomic formulas defined in Section 2, it is clear that $\phi$ does not  contain quantifiers.  
\end{proof}
 This corollary makes the connection between the set $A'$ and the notion of full elimination of quantifiers.   In fact, if $T$  is decomposable and does not accept full elimination of quantifiers then it is enough to add axioms to $T$ which  enable the elimination of all the quantifiers of the formulas of $A'$ to get a theory which accepts a full elimination of quantifiers. The sets $A''$ and $A'''$ are not concerned by this notion since in any decomposable theory $T$ the formulas of  $A''$ and $A'''$ accept full elimination of quantifiers due to their associated vectorial quantifiers:  $\exists!$ and $\exists^{\Psi(u)}_{\infty}$.  On the other hand, if $T$ is a decomposable theory which satisfies Corollary \ref{elimq} then we can  interest ourselves in getting the smallest subset $T^*$ of axioms of $T$, such that $T^*$ still accepts full elimination of quantifiers. For that it is enough to remove axiom by axiom from $T$ and check each time if the theory  still satisfies Corollary \ref{elimq}. This corollary shows  also the fact that  a decomposable theory $T$ does not  mean that $T$
accepts  full elimination of quantifiers. In fact, the theories of
infinite trees, finite trees and finite or infinite trees as
defined by M. Maher \cite{Maher} do not accept full elimination of
quantifiers but are decomposable and thus complete \cite{moi3}.

\subsection{Simple decomposable theories}
We present in this sub-section two examples of simple decomposable theories. The first one is a simple axiomatization of an infinite set of distinct individuals with an empty set of function and relation symbols. This theory denoted by $Eq$ can be seen as a small extension  of the Clark equational theory CET \cite{clark}, even if  according to our syntax the equality symbol is considered as a primitive logical symbol together with its usual properties (commutativity, transitivity ...). The second theory is the theory of additive  rational or real numbers with addition and subtraction. The goal of these examples is  to show the decomposability of  simple theories whose properties are well known and do not need addition of proofs.  An other example of a non-simple decomposable theory (finite or infinite trees) is given in Section \ref{arbre} with a detailed study of the properties of this theory.

Let us assume for all this sub-section  that the variables of $V$ are ordered by a strict linear dense order relation without endpoints denoted by $\succ$.
\subsection*{Equality theory}
Let $Eq$ be a theory together with an empty set of function and relation symbols and whose axioms is the infinite set of propositions of the following form
\begin{equation}\label{axioeq}
(1_n) \;\; \forall x_1...\forall x_n\exists y\, \neg(x_1=y)\wedge...\wedge\neg(x_n=y),
\end{equation}
where all the variables $x_1$...$x_n$ are distinct and $(n\neq 0)$. The form (\ref{axioeq}) is called \emph{diagram of axiom} and for each value of $n$ there exists \emph{an axiom} of $Eq$.  For example the following property is true in $Eq$:\[Eq\models\exists x\,\neg(x=y)\wedge\neg(x=z).\] The theory $Eq$ has as model an infinite set of distinct individuals.

Note that since $Eq$ has an empty set of function and relation symbols, then $AT=FL$ and thus  all the equations of $Eq$ are flat equations. Let $x$ and $y$ be two distinct variables. We call \emph{leader} of the equation $x=y$ the variable $x$. A conjunction $\alpha$ of flat formulas  is called $(\succ)$-\emph{solved} in $Eq$ if: (1) $\faux$ is not a sub-formula of $\alpha$,  (2)  if   $x=y$ is a sub-formula  of $\alpha$  then\footnote{Recall that $\succ$ is a strict linear dense order relation and thus $x\not\succ x$. In other terms $x=x$ is not ($\succ$)-solved.} $x\succ y$, (3) each equation of $\alpha$ has a distinct leader which does not occur in the other equations of $\alpha$.  

\begin{propriete}\label{theq1}
Every conjunction of flat formulas is equivalent in $Eq$ either to $\faux$ or to a $(\succ)$-solved conjunction of equations. 
 \end{propriete}
 Let $x$, $y$ and $z$ be variables such that $x\succ y\succ z$. The conjunction $x=x\wedge y=z$ is not $(\succ)$-solved because  in the equation $x=x$ we have $x\not\succ x$. By the same way, the conjunction $x=y\wedge y=z$ is not $(\succ)$-solved because $y$ is leader in $y=z$ and  occurs also in $x=y$.  The conjunctions $\vrai$ and $x=z\wedge  y=z$ are $(\succ)$-solved. The  computation of a possibly $(\succ)$-solved conjunction of equations from a conjunction of flat formulas in $Eq$ is evident\footnote{\[\begin{array}{l}(1)\; y=x\Longrightarrow x=y.\;\; (2)\; x=y\wedge x=z\Longrightarrow x=y\wedge z=y.\;\;(3)\;x=y\wedge z=x\Longrightarrow x=y\wedge z=y.\\ (4)\; \faux\wedge\alpha\Longrightarrow\faux.\;\;(5)\; x=x\Longrightarrow\vrai.\end{array}\] The rules (1), (2) and (3) are applied only if $x\succ y$.} and proceeds using the usual properties of the equality (commutativity, substitution, transitivity... ) and by replacing each formula of the form $x=x$ respectively $\alpha\wedge\faux$ by $\vrai$ respectively by $\faux$. 
\begin{propriete}\label{theq2}
Let $\alpha$ be a $(\succ)$-solved conjunction of equations. Let $\bar{x}$  be the vector of the leaders of the equations of $\alpha$. We have:
\begin{enumerate}
\item
$Eq\models\exists!\bar{x}\,\alpha$.
\item

For all $x\in V$ we have $Eq\models\exists^{\{\faux\}}_{\infty}x\,\vrai$.

\item
For all $x\in var(\alpha)$ we have $Eq\models\exists? x\,\alpha$.

\end{enumerate}
 \end{propriete}
 The first point holds because all the leaders of the equations of $\alpha$ are distinct and have one and only occurrence in $\alpha$. Thus, for every instantiation of the right hand sides of each equation, there exists one and only one value for the left hand sides and thus for the leaders.  The second point is a consequence of  the diagram of axiom (\ref{axioeq}) which states that for every finite set of distinct variables $x_1...x_n$ there exists a variable $y$ which is different from all the $x_i$. Thus, in each model of $Eq$ there exists an infinite set  of individuals. Thus according to  Definition \ref{infini}  we have 
$Eq\models\exists^{\{\faux\}}_{\infty}x\,\vrai$.  The third  point holds since in a ($\succ$)-solved conjunction of equations we have no formulas of the form $x=x$ (because $x\not\succ x$). Thus, using the properties of the equality for every model of $Eq$ and for every instantiation of the  variables of $var(\alpha)-\{x\}$ either there exists a unique solution of $x$ or there exists  a contradiction in the instantiations and thus there exists no values for $x$.
 
\begin{propriete}
The theory $Eq$ is decomposable.
\end{propriete}
\begin{proof}
 We show that $Eq$ satisfies the conditions of Definition \ref{decomp}. The sets $A$, $A'$, $A''$, $A'''$ and $\Psi(u)$ are chosen as follows:
 \begin{itemize}
 \item
 $A$ is the set $FL$.
 \item
 $A'$ is the set of formulas of the form $\exists\varepsilon\,\alpha'$ where $\alpha'$ is either a $(\succ)$-solved conjunction of equations  or the formula $\faux$.
\item
$A''$ is the set of formulas of the form $\exists\bar{x}''\,\vrai$.
\item
$A'''$ is the set of formulas of the form $\exists\bar{x}'''\alpha'''$ with $\alpha'''$ a $(\succ)$-solved conjunction of equations and $\bar{x}'''$  the vector of the leaders of the equations of $\alpha'''$.
\item $\Psi(u)=\{\faux\}$.
\end{itemize}
It is obvious that $FL$ is $Eq$-closed and $A'$, $A''$ and $A'''$ contain formulas  of the form $\exists\bar{x}\,\alpha$ with $\alpha\in FL$. 

 Let us show that $Eq$ satisfies the first condition of Definition \ref{decomp}. Let $\psi$ be any formula and  $\alpha\in FL$. Let $\bar{x}$ be a vector of variables. Let us choose an order $\succ$ such that the variables of $\bar{x}$ are greater than the free variables of $\exists\bar{x}\,\alpha$. According to Property \ref{theq1} two cases arise: 

- If the formula $\alpha$ is equivalent to $\faux$ in $Eq$, then  the formula $\exists\bar{x}\alpha\wedge\psi$ is equivalent in $Eq$ to a decomposed formula of the form 
 \[\exists \varepsilon\,\faux\wedge(
\exists\varepsilon\,\vrai\wedge(\exists\varepsilon\,\vrai\wedge
\psi)).
 \]
 
- If the formula $\alpha$ is equivalent in $Eq$ to a  ($\succ$)-solved conjunction $\beta$ of  equations, then let $X_l$ be the set of the variables of $\bar{x}$ which are leader in the equations of $\beta$ and let $X_n$ be the set of the variables of $\bar{x}$ which are not leader in the equations of $\beta$. The formula $\exists\bar{x}\alpha\wedge\psi$ is equivalent in $Eq$ to a decomposed formula of the form
\begin{equation}\label{x1}
\exists \bar{x}'\,\alpha'\wedge(
\exists\bar{x}''\,\alpha''\wedge(\exists\bar{x}'''\,\alpha'''\wedge
\psi)),
\end{equation}
with $\bar{x}'=\varepsilon$. The formula $\alpha'$ contains the conjunction of the equations of $\beta$ whose leaders do not belong to $X_l$. The vector $\bar{x}''$ contains the variables of $X_n$. The formula $\alpha''$ is the formula $\vrai$. The vector $\bar{x}'''$ contains the variables of $X_l$. The formula $\alpha'''$ is the conjunction of the equations of $\beta$ whose leaders belong to $X_l$. According to our construction it is clear that  $\exists\bar{x}'\alpha'\in A'$, $\exists\bar{x}''\alpha''\in A''$ and  $\exists\bar{x}'''\alpha\in A'''$.  Let us show that (\ref{x1}) and $\exists\bar{x}\alpha\wedge\psi$ are equivalent in $Eq$. Let $X$, ${X}'$, ${X}''$ and ${X}'''$ be the sets of the variables of the vectors\footnote{Of course if $\bar{x}=\varepsilon$ then $X=\emptyset$}  $\bar{x}$, $\bar{x}'$, $\bar{x}''$ and  $\bar{x}'''$. If $\alpha$ is equivalent to $\faux$ in $Eq$ then the equivalence of the decomposition is evident. Else $\beta$ is a ($\succ$)-solved conjunction of  equations and thus according to our construction we have:    ${X}={X}'\cup{X}''\cup{X}'''$, ${X}'\cap{X}''=\emptyset$, ${X}'\cap{X}'''=\emptyset$, ${X}''\cap{X}'''=\emptyset$, $X'=\emptyset$, for all $x''_i\in {X}''$ we have $x''_i\not\in var(\alpha')$ and for all $x'''_i\in{X}'''$ we have  $x'''_i\not\in var(\alpha'\wedge\alpha'')$. This is due to  the definition of the ($\succ$)-solved conjunction of flat formulas and the  order $\succ$ which has been chosen such that the quantified variables of $\exists\bar{x}\,\alpha$ are greater than the free variables of $\exists\bar{x}\,\alpha$. On the other hand, each equation in $\beta$ occurs  in $\alpha'\wedge\alpha''\wedge\alpha'''$  and each equation in $\alpha'\wedge\alpha''\wedge\alpha'''$ occurs in $\beta$ and thus $Eq\models\beta\leftrightarrow(\alpha'\wedge\alpha''\wedge\alpha''')$. We have shown that the vectorial quantifications  are coherent and the equivalence $Eq\models\beta\leftrightarrow\alpha'\wedge\alpha''\wedge\alpha'''$ holds. According to Property \ref{theq1} we have $Eq\models\alpha\leftrightarrow\beta$ and thus, the decomposition keeps the equivalence in $Eq$. Let us decompose for example \[\exists xyz\, v=w\wedge z=z\wedge z=x\wedge v=y .\]
Let us choose the order $\succ$ such that $x\succ y\succ z\succ v\succ w$. Note that the quantified variables are greater than the free variables. Let us now ($\succ$)-solve the conjunction $v=w\wedge z=z\wedge z=x\wedge v=y$. Thus the preceding formula is equivalent in $Eq$ to 
\[\exists xyz\, v=w\wedge  x=z\wedge y=w.\] We have $X_l=\{x,y\}$ and $X_n=\{z\}$. Thus, the preceding  formula is equivalent in $Eq$ to the following  decomposed formula
\[\exists\varepsilon\, v=w\wedge(\exists z\,\vrai\wedge(\exists xy\, x=z\wedge y=w)).\]

The theory $Eq$ satisfies the second  condition  of Definition \ref{decomp} according to the third point of Property \ref{theq2} and using the fact that $\bar{x}'=\varepsilon$. The theory $Eq$ satisfies the third  condition  of Definition \ref{decomp} according to the second point of Property \ref{theq2}. The theory $Eq$ satisfies the fourth  condition  of Definition \ref{decomp} according to the first point of Property \ref{theq2}. The theory $Eq$ satisfies the last condition  of Definition \ref{decomp} because $A'$ is of the form $\exists\varepsilon\,\alpha'$ where $\alpha'$ is either the formula $\faux$ or a $(\succ)$-solved conjunction of equations. Thus, if $\exists\varepsilon\,\alpha'$ has no free variables, then either $\alpha'=\vrai$ or $\alpha'=\faux$.
\end{proof}

Note that $Eq$ accepts full elimination of quantifiers. In fact Corollary \ref{elimq} illustrates this result since for all $\exists\bar{x}'\alpha'\in A'$ we have $\bar{x}'=\varepsilon$.

\subsection*{Additive rational or real numbers theory}
Let  $F=\{+,-,0,1\}$ be a set of function symbols of respective arities $2,1,0,0$. Let  $R=\emptyset$ be an empty set of relation symbols. Let  $Ra$ be the theory of  additive rational or real numbers together with  addition and subtraction. 
\begin{notation}\label{ahz}
Let  $a$ be a positive integer  and 
$t_1,..., t_n$ terms. We denote by:
\begin{itemize}
\item $Z$ the set of the integers.
\item $t_1+t_2$, the term
$+t_1t_2$. \item $t_1+t_2+t_3$, the term $+t_1(+t_2t_3)$. \item $0.t_1$, the term 0. \item
$-a.t_1$, the term $\underbrace{(-t_1) + \cdots + (-t_1)}_{a}$.
 \item $a.t_1$, the term $\underbrace{t_1
+ \cdots + t_1}_{a}$. \item $\sum_{i=1}^{n} t_i$, the term
$\overline{t_1+t_2+...+t_n}+0$, where $\overline{t_1+t_2+...+t_n}$
is the term $t_1+t_2+...+t_n$ in which we have removed all the
$t_i$'s which are equal to $0$. For $n=0$ the term $\sum_{i=1}^{n} t_i$ is reduced to the term $0$.
\end{itemize}
\end{notation}

The axiomatization of $Ra$  is the set of propositions of one of the  8 following forms: 
\[\begin{array}{lll}
&1 & \forall x\forall y \, x+y=y+x,\\
&2 & \forall x\forall y\forall z\,x+(y+z)=(x+y)+z,\\
&3 &\forall x\,x+0=x,\\
&4 &\forall x\, x+(-x)=0,\\
&5_n &\forall x \, n.x=0 \rightarrow x=0,\\
&6_n &\forall x \, \exists! y\, n.y=x,\\& 7 & \forall x\forall y\forall z\,(x=y)\leftrightarrow (x+z=y+z),\\&8 & \neg(0=1).
\end{array}\]
with $n$ an non-null integer. This theory has two usual models:  rational numbers $Q$ with addition and subtraction in $Q$ and real numbers $R$ with addition and subtraction in $R$.

We call \emph{block} every conjunction $\alpha$ of formulas of the
form:
 $\vrai$,  $\faux$,  
$\sum_{i=1}^{n} a_i.x_i=a_0.1$  with $x_1,...,x_n$ distinct variables and $a_i\in Z$ for all $i\in\{0,1,...,n\}$. We call leader of an equation of the form $\sum_{i=1}^{n} a_i.x_i=a_0.1$ the greatest variables $x_k$ ($k\in\{1,...,n\}$) according to the order $\succ$ such that $a_k\neq 0$. A block $\alpha$ is called  $(\succ)$-\emph{solved} in $Ra$ if (1) each equation of $\alpha$ has a distinct leader which does not occur in the other equations of $\alpha$ and (2) $\alpha$ does not contain sub-formulas of the form $0=a_0.1$ or $\faux$ with $a_0\in Z$. According to the axiomatization of $Ra$ we show easily that:

\begin{propriete}\label{reverse}
For all $k\in\{1,...,n\}$ we have:
\[Ra\models \sum_{i=1}^{n} a_i.x_i=a_0.1\leftrightarrow a_k.x_k=\sum_{i=1,i\neq k}^{n} (-a_i).x_i+a_0.1\]
\end{propriete}

\begin{propriete}\label{thra1}
Every block is equivalent in $Ra$ either to $\faux$ or to a $(\succ)$-solved block.
\end{propriete}
Let $x$, $y$ and $z$ be variables such that $x\succ y\succ z$. The block $2.x+y=(-1).1\wedge 2.z+y=2.1$ is not $(\succ)$-solved because $y$ is leader in the second equation and occurs also in the first one.  By the same way, the block $x+y=3.1\wedge 0=0.1$ is not $(\succ)$-solved because $0=0.1$ occurs in it.  The blocks  $\vrai$ and $x+2.z=4.1\wedge  3.y+2.z=3.1$ are $(\succ)$-solved. The  computation of a possibly $(\succ)$-solved block is evident\footnote{\[\begin{array}{l}
(1)\; 0=0.1\Longrightarrow \vrai.\;\;(2)\; 0=a_0.1\Longrightarrow\faux.\;\;(3)\;\faux\wedge\alpha\Longrightarrow\faux.\\
(4)\;\left[\begin{array}{@{}l@{}}
       \sum_{i=1}^{n} a_i.x_i=a_0.1\wedge\\\sum_{i=1}^{n}
       b_i.x_i=b_0.1
       \end{array}\right]
\Longrightarrow \left[\begin{array}{@{}ll@{}}
 \sum_{i=1}^{n} a_i.x_i=a_0.1\wedge\\
 \sum_{i=1}^{n} (b_ka_i-a_kb_i).x_i=(b_ka_0-a_kb_0).1
 \end{array}\right]. \end{array}\]In the rule (2) $a_0\neq 0$. In the rule (4) $x_k$ is the leader of the block  $\sum_{i=1}^{n} a_i.x_i=a_0.1$ and $b_k\neq 0$.}
and proceeds  using Property \ref{reverse} and a usual technique of substitution and simplification  by replacing each equation of the form $0=a_0.1$  by $\faux$ if  $a_0\neq 0$ and by $\vrai$ otherwise and each formula of the form $\faux\wedge\alpha$ by $\faux$.
\begin{propriete}\label{thra2}
Let $\alpha$ be a $(\succ)$-solved block and $\bar{x}$ be the vector of the leaders of the equations of $\alpha$. We have:
\begin{enumerate}
\item
$Ra\models\exists!\bar{x}\,\alpha$.
\item
For all $x\in V$ we have $Ra\models\exists^{\{\faux\}}_{\infty}x\,\vrai$.
\item
For all $x\in var(\alpha)$ we have $Ra\models\exists? x\,\alpha$.
\end{enumerate}
 \end{propriete}
 The first point holds because all the leaders are distinct and do not occur in the other equations.  Thus, if we transform each equation of the form  $\sum_{i=1}^{n} a_i.x_i=a_0.1$ using Property \ref{reverse} into a formula of the form $a_k.x_k=\sum_{i=1,i\neq k}^{n} (-a_i).x_i+a_0.1$ with $x_k$ the leader of this equation, then we get a conjunction of equations whose left hand sides are distinct and do not occur in the right hand sides. Thus, for each instantiation of the right hand sides of these equations there exists one and only value for the left hand sides and thus for the leaders according to axiom 6 of $Ra$.   The second point holds because according to  axiom 8 we have $Ra\models\neg(0=1)$ thus using  axiom 7 we have $Ra\models\neg(0+1=1+1)$. Then using  axiom 3 we get $Ra\models\neg(1=1+1)$. Thus using the transitivity of the equality we have $Ra\models\neg(0=1+1)$. If we repeat the preceding  steps $n$ times we get $n$ different individuals in all models of $Ra$. Thus for every model of $Ra$ there exists an infinite set of individuals. Thus according to  Definition \ref{infini}  we have 
$Ra\models\exists^{\{\faux\}}_{\infty}x\,\vrai$.  The third  point is evident according to the form of the  blocks and the definition of the ($\succ$)-solved block. 
 \begin{propriete}
The theory $Ra$ is decomposable.
\end{propriete}
\begin{proof}
 We show that $Ra$ satisfies the conditions of Definition \ref{decomp}. The sets $A$, $A'$, $A''$, $A'''$ and $\Psi(u)$ are chosen as follows:
 \begin{itemize}
 \item
 $A$ is the set of blocks.
 \item
 $A'$ is the set of formulas of the form $\exists\varepsilon\,\alpha'$ where $\alpha'$ is either a $(\succ)$-solved block or the formula $\faux$.
\item
$A''$ is the set of formulas of the form $\exists\bar{x}''\,\vrai$.
\item
$A'''$ is the set of formulas of the form $\exists\bar{x}'''\alpha'''$ with $\alpha'''$ a $(\succ)$-solved block and $\bar{x}'''$  the vector of the leaders of the equations of $\alpha'''$.
\item $\Psi(u)=\{\faux\}$.
\end{itemize}
Let us denote by $BL$ the set of the blocks. It is  clear that   $A'$, $A''$ and $A'''$ contain formulas of the form   $\exists\bar{x}\,\alpha$ with $\alpha\in BL$. Let us show that $BL$ is $Ra$-closed: (i) According to the definition of $BL$ we have $BL\subseteq AT$. (ii)  $BL$ is closed under conjunction. (iii)  Let  $\alpha$ be a flat formula. If $\alpha$ is the formula $\vrai$, $\faux$, $x=0$ or $x=1$ then it is a block\footnote{The formulas $x=0$ and $x=1$ are  blocks because the notations $1.x$,   $0.1$ and  $1.1$ denote the terms $x$, $0$ and $1$  according to Notation \ref{ahz}}. Else the following transformations transform $\alpha$ to a block
\[\begin{array}{lll}
x=y&\Longrightarrow& x+(-1).y=0.1 \\
x=-y&\Longrightarrow& x+y=0.1 \\
x=y+z&\Longrightarrow& x+(-1).y+(-1).z=0.1 \\
\end{array}
\]
From (i), (ii) and (iii)  $BL$ is $Ra$-closed.  Let us show that $Ra$ satisfies the first condition of Definition \ref{decomp}. Let  $\psi$ be any formula and $\alpha\in BL$. Let $\bar{x}$ be a vector of variables. Let us choose an order $\succ$ such that the variables of $\bar{x}$ are greater than the free variables of $\exists\bar{x}\,\alpha$. According to Property \ref{thra1} two cases arise: 

- If $\alpha$ is equivalent to $\faux$ in $Ra$, then the formula $\exists\bar{x}\alpha\wedge\psi$ is equivalent in $Ra$ to a decomposed formula of the form 
 \[\exists \varepsilon\,\faux\wedge(
\exists\varepsilon\,\vrai\wedge(\exists\varepsilon\,\vrai\wedge
\psi)).
 \]
 
- If $\alpha$ is equivalent in $T$ to a  ($\succ$)-solved block $\beta$, then let $X_l$ be the set of the variables of $\bar{x}$ which are leader in the equations of $\beta$ and let $X_n$ be the set of the variables  of $\bar{x}$  which are not leader in the equations of $\beta$. The formula $\exists\bar{x}\alpha\wedge\psi$ is equivalent in $T$ to a decomposed formula of the form
\begin{equation}\label{x2}
\exists \bar{x}'\,\alpha'\wedge(
\exists\bar{x}''\,\alpha''\wedge(\exists\bar{x}'''\,\alpha'''\wedge
\psi)),
\end{equation}
with $\bar{x}'=\varepsilon$. The formula $\alpha'$ contains the conjunction of the equations of $\beta$ whose leaders do not belong to $X_l$. The vector $\bar{x}''$ contains the variables of $X_n$. The formula $\alpha''$ is the formula $\vrai$. The vector $\bar{x}'''$ contains the variables of $X_l$. The formula $\alpha'''$ is the conjunction of the equations of $\beta$ whose leaders belong to $X_l$.  According to our construction it is clear that  $\exists\bar{x}'\alpha'\in A'$, $\exists\bar{x}''\alpha''\in A''$ and  $\exists\bar{x}'''\alpha\in A'''$.  Let us show that (\ref{x2}) and $\exists\bar{x}\alpha\wedge\psi$ are equivalent in $Ra$. Let $X$, ${X}'$, ${X}''$ and ${X}'''$ be the sets of the variables of the vectors $\bar{x}$,  $\bar{x}'$, $\bar{x}''$ and  $\bar{x}'''$. If $\alpha$ is equivalent to $\faux$ in $Ra$ then the equivalence of the decomposition is evident. Else  $\beta$ is a ($\succ$)-solved  block  and thus according to our construction we have:    ${X}={X}'\cup{X}''\cup{X}'''$, ${X}'\cap{X}''=\emptyset$, ${X}'\cap{X}'''=\emptyset$, ${X}''\cap{X}'''=\emptyset$, $X'=\emptyset$, for all $x''_i\in {X}''$ we have  $x''_i\not\in var(\alpha')$ and for all $x'''_i\in{X}'''$ we have  $x'''_i\not\in var(\alpha'\wedge\alpha'')$. This is due to the definition of ($\succ$)-solved blocks and  the order $\succ$ which has been chosen such that the quantified variables of $\exists\bar{x}\,\alpha$ are greater than the free variables of $\exists\bar{x}\,\alpha$. On the other hand,  each equation of $\beta$ occurs  in $\alpha'\wedge\alpha''\wedge\alpha'''$ and each equation in $\alpha'\wedge\alpha''\wedge\alpha'''$  occurs in $\beta$ and thus $Ra\models\beta\leftrightarrow(\alpha'\wedge\alpha''\wedge\alpha''')$. We have shown that the vectorial quantifications  are coherent and the equivalence $Ra\models\beta\leftrightarrow\alpha'\wedge\alpha''\wedge\alpha'''$ holds. According to Property \ref{thra1} we have $Ra\models\alpha\leftrightarrow\beta$ and thus, the decomposition keeps the equivalence in $Ra$. Let us decompose for example \[\exists xyz\, 2.v+w=3.1\wedge v+x=2.1\wedge v+x+2.z=4.1 \]
Let us choose the order $\succ$ such that $x\succ y\succ z\succ v\succ w$. Note that the quantified variables are greater than the free variables.  Let us now ($\succ$)-solve the block $2.v+w=3.1\wedge v+x=2.1\wedge v+x+2.z=4.1$. Thus the preceding formula is equivalent in $Ra$ to 
\[\exists xyz\, 2.v+w=3.1\wedge  2.x+(-1).w=1\wedge z=1\] We have $X_l=\{x,z\}$ and $X_n=\{y\}$  thus the preceding formula is equivalent in $Ra$ to the following  decomposed formula
\[\exists\varepsilon\, 2.v+w=3.1\wedge(\exists y\,\vrai\wedge(\exists xz\, 2.x+(-1).w=1\wedge z=1)).\]

The theory $Ra$ satisfies the second  condition  of Definition \ref{decomp} according to the third point of Property \ref{thra2} and using the fact that $\bar{x}'=\varepsilon$. The theory $Ra$ satisfies the third  condition  of Definition \ref{decomp} according to the second point of Property \ref{thra2}. The theory $Ra$ satisfies the fourth  condition  of Definition \ref{decomp} according to the first point of Property \ref{thra2}. The theory $Ra$ satisfies the last condition  of Definition \ref{decomp} because $A'$ is of the form $\exists\varepsilon\,\alpha'$ where $\alpha'$ is either a ($\succ$)-solved block or the formula $\faux$. Thus, if $\alpha'$ does not contain  free variables then  according to the definition of the $(\succ)$-solved blocks $\alpha'$ does not contain formulas of the form $0=a_01$ and thus  $\alpha'$ is either the formula $\vrai$ or the formula $\faux$. 
\end{proof}

Note that $Ra$ accepts full elimination of quantifiers. In fact Corollary \ref{elimq} illustrates this result since for all $\exists\bar{x}'\alpha'\in A'$ we have $\bar{x}'=\varepsilon$.

\section{A general algorithm for solving first-order formulas in a decomposable theory $T$}
Let $T$ be a decomposable theory together with its set of function symbols $F$ and its set of relation symbols $R$. The sets $\Psi(u)$, $A$, $A'$, $A''$
and $A'''$ are now known and fixed.
\subsection{Normalized  formula}
\begin{definition1} A normalized formula $\varphi$ of depth $d\geq 1$ is a formula of the form
\begin{equation}\label{norm}
\neg(\exists\bar{x}\,\alpha\wedge\bigwedge_{i\in I} \varphi_i),
\end{equation}
 with $I$ a finite (possibly empty) set, $\alpha\in FL$
 and the $\varphi_i's$ are normalized formulas of depth $d_i$ with
 $d=1+\max\{0,d_1,...,d_n\}$ and all the quantified variables of $\varphi$ have
 distinct names and different from the names of the free
 variables.
 \end{definition1}

 \begin{example}
 Let $f$ and $g$ be two $1$-ary function symbols which belong to $F$. The formula 
 \[\neg\left[\exists\varepsilon\vrai\wedge\left[\begin{array}{l} \neg(\exists x\,y=fx\wedge x=y\wedge\neg(\exists\varepsilon\,y=gx))\wedge\\\neg(\exists\varepsilon\,x=z)\end{array}\right]\right]\]
is a  normalized formula of depth equals to three. The formulas $\neg(\exists\varepsilon\,\vrai)$ and $\neg(\exists\varepsilon\,\faux)$ are two   normalized formulas of depth 1.  The smallest value of a depth of a normalized formula is 1. Normalized formulas of depth 0 are not defined and do not exist.
 \end{example}

\begin{propriete}\label{norm11} 
Every formula $\varphi$ is equivalent in $T$ to a wnfv  normalized formula of depth $d\geq 1$.
\end{propriete}
\begin{proof}
 It is easy to transform any formula to a wnfv normalized formula, it is enough for example to follow the followings steps: \begin{enumerate}\item 
Introduce a supplement of equations and existentially quantified
variables to transform the conjunctions of atomic formulas into
conjunctions of flat formulas. \item Express all the
quantifiers, constants and logical connectors using only the logical symbols $\neg$, $\wedge$
and $\exists$. This can be done using the following transformations\footnote{These equivalences are true in the empty theory and thus in any theory $T$. } of sub-formulas:
\[\begin{array}{lll}
(\varphi\vee\phi)&\Longrightarrow& \neg(\neg\varphi\wedge\neg\phi),\\

(\varphi\rightarrow\phi)&\Longrightarrow& \neg(\varphi\wedge\neg\phi),\\

(\varphi\leftrightarrow\phi)&\Longrightarrow& (\neg(\varphi\wedge\neg\phi)\wedge\neg(\phi\wedge\neg\varphi)),\\

(\forall x\,\varphi)&\Longrightarrow& \neg(\exists x\,\neg\varphi).\end{array}
\]

 \item If the formula $\varphi$ obtained
does not start with the logical symbol $\neg$, then  replace it by $\neg(\exists\varepsilon\,\vrai\wedge\neg \varphi)$.
\item  Name the quantified variables by distinct names and
different from the names of the free variables. \item Lift the
quantifier before the conjunction, i.e.
$\varphi\wedge(\exists\bar{x}\,\psi)$ or  $(\exists\bar{x}\,\psi)\wedge\varphi$, becomes
$\exists\bar{x}\,\varphi\wedge \psi$ because the free variables
of $\varphi$ are distinct from those of $\bar{x}$.
\item Group the quantified variables into a vectorial quantifier, i.e. $\exists\bar{x}(\exists\bar{y}\,\varphi)$ or $\exists\bar{x}\exists\bar{y}\,\varphi$ becomes $\exists\overline{xy}\,\varphi$.
\item Insert empty vectors and formulas of the form $\vrai$ to get the  normalized form using the following transformations of sub-formulas: 
\begin{equation}\label{regle1}
\neg(\bigwedge_{i\in I}\neg\varphi_i)\Longrightarrow \neg(\exists\varepsilon\,\vrai\wedge\bigwedge_{i\in I}\neg\varphi_i),\end{equation}\begin{equation}\label{regle2} \neg(\alpha\wedge\bigwedge_{i\in I}\neg\varphi_i)\Longrightarrow \neg(\exists\varepsilon\,\alpha\wedge\bigwedge_{i\in I}\neg\varphi_i),\end{equation}
\begin{equation}\label{regle3}
\neg(\exists\bar{x}\,\bigwedge_{j\in J}\neg\varphi_j) \Longrightarrow \neg(\exists\bar{x}\,\vrai\wedge\bigwedge_{j\in J}\neg\varphi_j).
\end{equation}
with $\alpha\in FL$, $I$ a finite (possibly empty) set and $J$ a finite non-empty set.
\end{enumerate}  If the starting
formula does not contain the logical symbol $\leftrightarrow$ then this
transformation will be linear, i.e. there exists a constant $k$
such that $n_2\leq kn_1$, where $n_1$ is the size of the starting
formula and $n_2$ the size of the normalized formula. We show easily by contradiction that the final formula obtained after application of these steps is normalized. 
\end{proof}

\begin{example}
Let $f$ be a $2$-ary function symbol which  belongs to $F$. Let us apply the preceding steps to transform the following formula into a normalized formula which is equivalent in $T$: \[(fuv=fwu\wedge(\exists x\, u=x))\vee(\exists u\,\forall w\, u=fvw).\]
Note that the  formula does not start with $\neg$ and the variables $u$ and $w$ are free in $fuv=fwu\wedge(\exists x\, u=x)$ and bound in $\exists u\,\forall w\, u=fvw$. $\\$Step 1: Let us first transform the equations into flat equations. The preceding formula is equivalent in $T$ to 
 \begin{equation}\label{ayay1}(\exists u_1\, u_1=fuv\wedge u_1=fwu\wedge(\exists x\, u=x))\vee(\exists u\,\forall w\, u=fvw).\end{equation}
Step 2: Let us now express the quantifier $\forall$ using $\neg$, $\wedge$ and $\exists$. Thus, the formula  (\ref{ayay1}) is equivalent in $T$ to

 \[(\exists u_1\, u_1=fuv\wedge u_1=fwu\wedge(\exists x\, u=x))\vee(\exists u\,\neg(\exists w\, \neg(u=fvw))).\]
Let us also express the logical symbol $\vee$ using $\neg$, $\wedge$ and $\exists$. Thus, the preceding formula is equivalent in $T$ to 
\begin{equation}\label{ayay2} \neg (\neg(\exists u_1\, u_1=fuv\wedge u_1=fwu\wedge(\exists x \,u=x))\wedge\\\neg(\exists u\,\neg(\exists w\, \neg(u=fvw)))).\end{equation}
Step 3: The formula starts with $\neg$, then we move to Step 4.$\\$
Step 4: The occurrences of the quantified variables $u$ and $w$ in $(\exists u\,\neg(\exists w\, \neg(u=fvw)))$ must be renamed. Thus, the formula (\ref{ayay2}) is equivalent in $T$ to 
\[\neg (\neg(\exists u_1\, u_1=fuv\wedge u_1=fwu\wedge(\exists x \,u=x))\wedge\neg(\exists u_2\,\neg(\exists w_1\, \neg(u_2=fvw_1)))).\]
Step 5: By lifting the existential quantifier  $\exists x$, the preceding formula is equivalent in $T$ to 
\[\neg (\neg(\exists u_1\,\exists x\, u_1=fuv\wedge u_1=fwu\wedge u=x)\wedge\neg(\exists u_2\,\neg(\exists w_1\, \neg(u_2=fvw_1)))).\]
Step 6: Let us group the two quantified variables $x$ and $u_1$ into a vectorial quantifier. Thus, the preceding formula is equivalent in $T$ to 

\[\neg (\neg(\exists u_1x\, u_1=fuv\wedge u_1=fwu\wedge u=x)\wedge\neg(\exists u_2\,\neg(\exists w_1\, \neg(u_2=fvw_1)))).\]
Step 7: Let us introduce empty vectors of variables and formulas of the form $\vrai$ to get the  normalized formula. According to the rule (\ref{regle1}), the preceding formula is equivalent in $T$ to 
\[\neg \left[\exists\varepsilon\,\vrai\wedge\left[\begin{array}{l} \neg(\exists u_1x\, u_1=fuv\wedge u_1=fwu\wedge u=x)\wedge\\\neg(\exists u_2\,\neg(\exists w_1\, \neg(u_2=fvw_1)))\end{array}\right]\right],\]
which  using the rule  (\ref{regle2}) with $I=\emptyset$   is equivalent in $T$ to 
\[\neg \left[\exists\varepsilon\,\vrai\wedge\left[\begin{array}{l} \neg(\exists u_1x\, u_1=fuv\wedge u_1=fwu\wedge u=x)\wedge\\\neg(\exists u_2\,\neg(\exists w_1\, \neg(\exists\varepsilon\, u_2=fvw_1)))\end{array}\right]\right],\]
which   using the rule (\ref{regle3})  is equivalent in $T$ to  
\[\neg \left[\exists\varepsilon\,\vrai\wedge\left[\begin{array}{l} \neg(\exists u_1x\, u_1=fuv\wedge u_1=fwu\wedge u=x)\wedge\\\neg(\exists u_2\,\vrai\wedge\neg(\exists w_1\,\vrai\wedge \neg(\exists\varepsilon\, u_2=fvw_1)))\end{array}\right]\right].\]
This is a  normalized formula of depth 4.

\end{example}

\subsection{Working formula}\label{hihiho1} 
 \begin{definition1}
 \label{work} A \emph{working formula} $\varphi$ of depth $d\geq 1$ is a  formula of the form
\begin{equation}\label{aller1}
\neg(\exists\bar{x}\,\alpha\wedge\bigwedge_{i\in I} \varphi_i),
\end{equation}
 with $I$ a finite (possibly empty) set, $\alpha\in A$
 and the $\varphi_i's$ are working formulas of depth $d_i$ with
 $d=1+\max\{0,d_1,...,d_n\}$ and all the quantified variables of $\varphi$ have
 distinct names and different from the names of the free
 variables. Working formulas of depth 0 are not defined and do not exist.

 \end{definition1}

\begin{propriete}\label{hihiho}
Every formula is equivalent in $T$ to a wnfv working formula.
\end{propriete}
\begin{proof}
Let $\varphi$ be a formula. According to Property \ref{norm11}, $\varphi$ is equivalent in $T$ to a wnfv normalized formula $\phi$ of the form 
\begin{equation}\label{alez1}
\neg(\exists\bar{x}\,\alpha\wedge\bigwedge_{i\in I} \varphi_i),
\end{equation}
with $\alpha\in FL$, $I$ a finite possibly empty set  and all the $\varphi_i$ are normalized formulas.  Let us show by recurrence on the depth $d$ of (\ref{alez1}) that the formula (\ref{alez1}) is equivalent in $T$ to a working formula. 

(1) Let us show first that the recurrence is true for  $d=1$, i.e. every   normalized formula of the form $\neg(\exists\bar{x}\,\alpha)$ with $\alpha\in FL$ is equivalent in $T$ to a working formula. Since $T$ is decomposable then according to Definition \ref{decomp} the set $A$ is $T$-closed, i.e. (i) $A\subseteq AT$, (ii) $A$ is closed under conjunction and (iii) every flat formula is equivalent in $T$ to a formula which belongs to $A$. Since $\alpha\in FL$, then according to (iii) $\alpha$ is equivalent in $T$ to a conjunction $\beta$ of elements of $A$. According to (ii) $\beta$ belongs to $A$. Thus, the formula $\neg(\exists\bar{x}\,\alpha)$ is equivalent in $T$ to $\neg(\exists\bar{x}\,\beta)$ with $\beta\in A$ which is a working formula of depth 1. 

(2) Let us suppose now that the recurrence is true for $d\leq n$ and let us show that it is true for $d=n+1$. Let  \begin{equation}\neg(\exists\bar{x}\,\alpha\wedge\bigwedge_{i\in I} \varphi_i),
\end{equation}
 be a normalized formula of depth $n+1$ with $\alpha\in FL$ and all the $\varphi_i$ are normalized formulas of depth $d_i\leq n$. According to the hypothesis of recurrence the preceding formula is equivalent in $T$ to a formula of the form 
 \begin{equation}\label{alez2}\neg(\exists\bar{x}\,\alpha\wedge\bigwedge_{i\in I} \varphi_i),
\end{equation}
with $\alpha\in FL$ and all the $\varphi_i$ are working formulas. Since $T$ is decomposable then according to Definition \ref{decomp} the set $A$ is $T$-closed, i.e. (i) $A\subseteq AT$, (ii) $A$ is closed under conjunction and (iii) every flat formula is equivalent in $T$ to a formula which belongs to $A$. Since $\alpha\in FL$, then according to (iii) $\alpha$ is equivalent in $T$ to a conjunction $\beta$ of elements of $A$. According to (ii) $\beta$ belongs to $A$. Thus, the formula (\ref{alez2}) is equivalent in $T$ to 
\[\neg(\exists\bar{x}\,\beta\wedge\bigwedge_{i\in I} \varphi_i),
\]
with $\beta\in A$ and all the $\varphi_i$ are working formulas. The preceding formula is a working formula. From (1) and (2) our recurrence is true.
\end{proof}

\begin{example}
In the theory $Ra$ of additive rational numbers, the formula 
\[\neg \left[\exists\varepsilon\,\vrai\wedge\left[\begin{array}{l} \neg(\exists x\,  y=-z\wedge z=x+y)\wedge\\\neg(\exists \,\vrai\wedge\neg(\exists w\,\vrai\wedge \neg(\exists\varepsilon\, z=w)))\end{array}\right]\right],\]
is a normalized formula of depth 4 which is equivalent in $Ra$ to the following working formula
\[\neg \left[\exists\varepsilon\,\vrai\wedge\left[\begin{array}{l} \neg(\exists x\,  y+z=0.1\wedge z+(-1).x+(-1).y=0.1)\wedge\\\neg(\exists \,\vrai\wedge\neg(\exists w\,\vrai\wedge \neg(\exists\varepsilon\, z+(-1).w=0.1)))\end{array}\right]\right].\]

\end{example}

\begin{definition1}\label{defs}
A \emph{solved} formula is a working formula of the form
\begin{equation}\label{res}
\neg(\exists\bar{x}'\,\alpha'\wedge\bigwedge_{i\in
I}\neg(\exists\bar{y}'_i\,\beta'_i)),
\end{equation}
where $I$ is a finite (possibly empty) set, $\exists\bar{x}'\alpha'\in
A'$, $\exists\bar{y}'_i\beta_i'\in A'$ for all $i\in I$, $\alpha'$ is different from
the formula $\faux$ and all the $\beta'_i$ are different from the
formulas $\vrai$ and $\faux$.
\end{definition1}

\begin{propriete}\label{uniq}
Let $\varphi$ be a conjunction of solved formulas without free
variables. The conjunction $\varphi$ is either the formula
$\neg\vrai$ or the formula $\vrai$.
\end{propriete}

\begin{proof}
Recall first that we
write $\bigwedge_{i\in I}\varphi_i$, and call \emph{conjunction}
each formula of the form
$\varphi_{i_1}\wedge\varphi_{i_2}\wedge...\wedge\varphi_{i_n}\wedge\vrai$.
Let $\varphi$ be  a conjunction of solved formulas without free
variables. According to Definition \ref{defs}, $\varphi$ is of the
form
\begin{equation}\label{mdr} (\bigwedge_{i\in I}\neg
(\exists\bar{x}'_i\alpha'_i\wedge\bigwedge_{j\in
J_i}\neg(\exists\bar{y}'_{ij}\beta'_{ij})))\wedge\vrai\end{equation}
with
\begin{enumerate} \item $I$ a finite (possibly empty) set,
\item $(\exists\bar{x}'_i\alpha'_i)\in A'$ for all $i\in I$, \item
$(\exists\bar{y}'_{ij}\beta'_{ij})\in A'$ for all $i\in I$ and $j\in J_i$, \item  $\alpha'_i$
different from $\faux$ for all $i\in I$, \item $\beta'_{ij}$ different from $\vrai$ and $\faux$ for all $i\in I$ and $j\in J_i$. \end{enumerate} Since these solved formulas don't
have free variables and since $T$ is a decomposable theory then
according to the fifth point of Definition \ref{decomp} of a
decomposable theory and the conditions 2 and 3 of (\ref{mdr}) we
have:$\\$ (*) each formula $\exists\bar{x}'_i\alpha'_i$ and each
formula $\exists\bar{y}'_{ij}\beta'_{ij}$ is either the formula
$\exists\varepsilon\vrai$ or $\exists\varepsilon\faux.\\$
According to (*) and the condition 5 of (\ref{mdr}),  all the sets
$J_i$ must be empty, thus $\varphi$ is of the form
\begin{equation}\label{mdr2} (\bigwedge_{i\in I}\neg
(\exists\bar{x}'_i\alpha'_i))\wedge\vrai\end{equation} According
to (*) and (\ref{mdr2}), the formula $\varphi$ is of the form
\[(\bigwedge_{i\in
I'}\neg(\exists\varepsilon\faux))\wedge(\bigwedge_{j\in
I-I'}\neg(\exists\varepsilon\vrai))\wedge\vrai\] According to the
condition 4 of (\ref{mdr}), the set $I'$ must be empty and thus
$\varphi$ is of the form
\[(\bigwedge_{i\in
I}\neg(\exists\varepsilon\vrai))\wedge\vrai\] If  $I=\emptyset$
then $\varphi$ is the formula $\vrai$. Else, according to our assumptions, we do not distinguish two formulas which
can be made equal using the following transformations of 
sub-formulas:
\[\begin{array}{@{}c@{}}\varphi\wedge\varphi\Longrightarrow\varphi,\;\;\varphi\wedge\psi\Longrightarrow\psi\wedge\varphi,\;\;(\varphi\wedge\psi)\wedge\phi\Longrightarrow\varphi\wedge(\psi\wedge\phi),\\\varphi\wedge\vrai\Longrightarrow\varphi,\;\;\varphi\vee\faux\Longrightarrow\varphi.\end{array}\]
Thus $\varphi$ is the formula
\[\neg\vrai\]
\end{proof}

\begin{propriete}\label{bool}
Every solved formula is equivalent in $T$ to a wnfv Boolean
combination of elements of $A'$.
\end{propriete}
\begin{proof}
Let $\varphi$ be a solved formula. According to Definition
\ref{defs}, the formula $\varphi$ is of the form \[
\neg(\exists\bar{x}'\,\alpha'\wedge\bigwedge_{i\in
I}\neg(\exists\bar{y}'_i\,\beta'_i)),
\]
with $\exists\bar{x}'\alpha'\in A'$ and
$\exists\bar{y}'_i\beta_i'\in A'$ for all $i\in I$. Since
$\exists\bar{x}'\alpha'\in A'$ then according to Definition
\ref{decomp} we have $T\models\exists?\bar{x}'\alpha'$ and thus according
to Corollary \ref{r2}, the preceding formula is equivalent in $T$ to the
following wnfv formula
\[
\neg((\exists\bar{x}'\,\alpha')\wedge\bigwedge_{i\in
I}\neg(\exists\bar{x}'\,\alpha'\wedge(\exists\bar{y}'_i\,\beta'_i))).
\]
According to the definition of working formula, all the quantified variables have  distinct names and different from the names of the
free  variables, thus the preceding formula is equivalent in $T$
to the wnfv formula
\[
\neg((\exists\bar{x}'\,\alpha')\wedge\bigwedge_{i\in
I}\neg(\exists\bar{x}'\bar{y}'_i\,\alpha'\wedge\beta'_i)).
\]
Since $\exists\bar{x}'\alpha'\in A'$ and
$\exists\bar{y}'_i\beta_i'\in A'$ for all $i\in I$, then $\alpha'\in A$ and
$\beta'_i\in A$. Since $A$ is $T$-closed then it is closed under conjunction and thus $\alpha'\wedge\beta'_i\in A$ for all $i\in I$. According
to Property \ref{dec2} the preceding  formula is equivalent in $T$
to a wnfv formula of the form
\[
\neg((\exists\bar{x}'\,\alpha')\wedge\bigwedge_{i\in
I}\neg(\exists\bar{z}'_i\,\delta'_i)),
\]
with $\exists\bar{x}'\alpha'\in A'$ and
$\exists\bar{z}'_i\delta_i'\in A'$ for all $i\in I$. Which is finally equivalent in
$T$ to
\[
(\neg(\exists\bar{x}'\,\alpha'))\vee\bigvee_{i\in
I}(\exists\bar{z}'_i\,\delta'_i).
\]

\end{proof}

\subsection{The rewriting rules}

We present now the rewriting rules which transform a working
formula $\varphi$ of any depth $d$ into a wnfv conjunction $\phi$ of solved formulas
which is equivalent to $\varphi$ in $T$. To apply the rule $p_1\Longrightarrow p_2$ to
the working formula $p$ means to replace in $p$, a sub-formula
$p_1$ by the formula $p_2$, by considering that the connector
$\wedge$ is associative and commutative.

\[\begin{array}{llll}

(1)& \neg\left[\begin{array}{l}
\exists\bar{x}\,\alpha\wedge\varphi\wedge\\[2mm]
\neg(\exists\bar{y}\,\vrai)
\end{array}\right]&\Longrightarrow &\;\vrai\\[8mm]

(2)& \neg\left[\begin{array}{l}
\exists\bar{x}\,\faux\wedge\varphi\\[2mm]
\end{array}\right]&\Longrightarrow &\;\vrai\\[7mm]

(3)\;\;& \neg\left[\begin{array}{l}
\exists\bar{x}\,\alpha\wedge\\[2mm]
\bigwedge_{i\in I}\neg(\exists\bar{y}_i\, \beta_i)
\end{array}\right]&\Longrightarrow \;\;\;\;\;&\neg\left[\begin{array}{l}
\exists\bar{x}'\bar{x}''\,\alpha'\wedge\alpha''\wedge\\[2mm]
\bigwedge_{i\in I}\neg(\exists\bar{x}'''\bar{y}_i\, \alpha'''\wedge\beta_i)^*\end{array}\right]\\[9mm]

(4)& \neg\left[\begin{array}{l}
\exists\bar{x}\,\alpha\wedge\\[2mm]
\bigwedge_{i\in I}\neg(\exists\bar{y}'_i\, \beta'_i)
\end{array}\right]&\Longrightarrow &\neg\left[\begin{array}{l}
\exists\bar{x}'\,\alpha'\wedge\\[2mm]
\bigwedge_{i\in I'}\neg(\exists\bar{y}'_i\,\beta'_i)\end{array}\right]\\[9mm]

(5)& \neg\left[\begin{array}{l} \exists\bar{x}\,\alpha\wedge
\varphi\wedge\\[2mm] \neg\left[\begin{array}{l} \exists\bar{y}'\,
\beta'\wedge\\[2mm]\bigwedge_{i\in I}\neg(\exists\bar{z}'_i\,\delta'_i)
\end{array}\right]\end{array}\right]&\Longrightarrow \;\;\;\;\;&\;\left[\begin{array}{l}
\neg(\exists\bar{x}\,\alpha\wedge\varphi\wedge\neg(\exists\bar{y}'\,\beta'))\wedge\\[2mm]\bigwedge_{i\in I}\neg(\exists\bar{x}\bar{y}'\bar{z}'_i\,\alpha\wedge\beta'\wedge \delta'_i\wedge\varphi)^*\\
\end{array}\right]\\[7mm]
\end{array}\]
with $\alpha\in A$, $\varphi$ a conjunction of working formulas
and $I$ a finite (possibly empty) set. In the rule (3),
 the formula $\exists\bar{x}\,\alpha$ is
equivalent in $T$ to a decomposed formula of the form
$\exists\bar{x}'\,\alpha'\wedge(\exists\bar{x}''\,\alpha''\wedge(\exists\bar{x}'''\,\alpha'''))$
with $\exists\bar{x}'\,\alpha'\in A'$,
$\exists\bar{x}''\,\alpha''\in A''$,
$\exists\bar{x}'''\,\alpha'''\in A'''$ and
$\exists\bar{x}'''\,\alpha'''$ different from
$\exists\varepsilon\,\vrai$. All the $\beta_i$'s belong to $A$. The formula $(\exists\bar{x}'''\bar{y}_i\, \alpha'''\wedge\beta_i)^*$ is the formula $(\exists\bar{x}'''\bar{y}_i\, \alpha'''\wedge\beta_i)$ in which we have renamed the variables of $\bar{x}'''$ by distinct names and different from the names of the free variables. In
the rule (4), the formula $\exists\bar{x}\,\alpha$ is not an
element of $A'$ and is equivalent in $T$ to a decomposed formula
of the form
$\exists\bar{x}'\,\alpha'\wedge(\exists\bar{x}''\,\alpha''\wedge(\exists\varepsilon\,\vrai))$
with $\exists\bar{x}'\,\alpha'\in A'$ and
$\exists\bar{x}''\,\alpha''\in A''$.  Each formula
$\exists\bar{y}'_i\,\beta'_i$ is an element of $A'$. $I'$ is the
set of the $i\in I$ such that $\exists\bar{y}'_i\beta'_i$ does not
have free occurrences of any variable of $\bar{x}''$. In the rule
(5), $I\neq\emptyset$, $\exists\bar{y}'\,\beta'\in A'$ and
$\exists\bar{z}'_i\,\delta'_i\in A'$ for all $i\in I$. The formula
$(\exists\bar{x}\bar{y}'\bar{z}'_i\,\alpha\wedge\beta'\wedge
\delta'_i\wedge\varphi)^*$ is the formula
$(\exists\bar{x}\bar{y}'\bar{z}'_i\,\alpha\wedge\beta'\wedge
\delta'_i\wedge\varphi)$ in which we have renamed the variables of
$\bar{x}$ and $\bar{y}'$ by distinct names and different from the
names of the free variables.

\begin{propriete}\label{gros}
Every repeated application of the preceding rewriting rules on
any working formula $\varphi$, terminates and produces  a
wnfv conjunction $\phi$ of solved formulas which is equivalent to $\varphi$
in $T$.
\end{propriete}
$\\$ {\emph{Proof, first part:}} The application of the rewriting
rules terminates. Let us consider the 3-tuple $(n_1,n_2,n_3)$
where the $n_i$'s are the following positive integers: $\\$\begin
{itemize}\item $n_1=\alpha(p)$, where the function $\alpha$ is
defined as follows: \begin{itemize} \item $\alpha(\vrai)=0$, \item
$\alpha(\neg(\exists\bar{x}\,a\wedge\varphi))=2^{\alpha(\varphi)}$,
\item $\alpha(\bigwedge_{i\in I}\varphi_i)=\sum_{i\in
I}\alpha(\varphi_i),$
\end{itemize}
with $a\in A$, $\varphi$ a conjunction of working formulas and
the $\varphi_i$'s working formulas. Note that if $\alpha(p_2) <
\alpha(p_1)$ then $\alpha(p[p_2]) < \alpha(p)$ where $p[p_2]$ is
the formula obtained from $p$ when we replace the occurrence of
the formula $p_1$ in $p$ by $p_2$. This function has been introduced in \cite{vo14} and \cite{dao2} to show the non-elementary  complexity of all algorithms solving propositions in the theory of finite or infinite trees. It has also  the property to decrease if the depth of the working formula decreases after application of  distributions as it is done in our rule (5). $\\$

\item $n_2=\beta(p)$, where the function $\beta$ is defined as
follows: \begin{itemize} \item $\beta(\vrai)=0$, \item
$\beta(\neg(\exists\bar{x}\,a\wedge\bigwedge_{i\in I}
\varphi_i))=\left\{\begin{array}{l} 4^{1+\sum_{i\in I}
\beta(\varphi_i)}$ if $
\exists\bar{x}'''\alpha'''\neq\exists\varepsilon\vrai,\\[1mm]1+\sum_{i\in I}\beta(\varphi_i)$
if $ \exists\bar{x}'''\alpha'''=\exists\varepsilon\vrai
\end{array}\right\}\\$
with the $\varphi_i$'s working formulas and
$T\models(\exists\bar{x}\alpha)\leftrightarrow(\exists\bar{x}'\alpha'\wedge(\exists\bar{x}''\alpha''\wedge(\exists\bar{x}'''\alpha''')))$.
\end{itemize} 
We show  that: \[\beta(\neg(\exists\bar{x}\,\alpha\wedge\bigwedge_{i\in I}\neg(\exists\bar{y}_i\,\lambda_i)))>\beta(\neg(\exists\bar{z}\delta\wedge\bigwedge_{i\in I}\neg(\exists w_i\,\gamma_i)))\] where  $I$ is a finite possibly empty set, the formula $\exists\bar{x}\,\alpha$ is equivalent in $T$ to a decomposed formula of the form $\exists\bar{x}'\alpha'\wedge(\exists\bar{x}''\alpha''\wedge(\exists\bar{x}'''\,\alpha'''))$ with $\exists\bar{x}'''\,\alpha'''\neq\exists\varepsilon\,\vrai$, the formula $\exists\bar{z}\,\delta$ is equivalent in $T$ to a decomposed formula of the form $\exists\bar{z}'\,\delta'\wedge(\exists\bar{z}''\,\delta''\wedge(\exists\varepsilon\,\vrai))$ and all the $\lambda_i$ and $\gamma_i$  belong to $A$ and have no particular  conditions.

 \item $n_3$ is the number of sub-formulas of the form
$\neg(\exists\bar{x}\alpha\wedge\varphi)$ with
$\exists\bar{x}\alpha\not\in A'$ and $\varphi$ a conjunction of
working formulas. \end{itemize}
For each rule, there exists an integer $i$ such that the application of
this rule decreases or does not change the values of the $n_j$'s,
with $1\leq j<i$, and decreases the value of $n_i$. This integer $i$ is
equal to:  1 for the rules (1), (2) and (5), 2 for the rule (3)
and 3 for the rule (4). To each sequence of formulas obtained by a
finite application of the preceding rewriting rules, we can
associate a series of 3-tuples $(n_1,n_2,n_3)$ which is strictly
decreasing in the lexicographic order. Since the $n_i$'s are
positive integers, they cannot be negative, thus this series of
3-tuples is a finite series and the application of the rewriting
rules terminates.

$\\[3mm]${\emph{Proof, second part:}} Let us show now that for each rule of the form
$p\Longrightarrow p'$ we have $T\models p\leftrightarrow p'$ and
the formula $p'$ remains a conjunction of working formulas. It
is clear that the rules (1) and (2) are correct.
\subsubsection*{Correctness of the rule (3):}
\[\neg\left[\begin{array}{l}
\exists\bar{x}\,\alpha\wedge\\[2mm]
\bigwedge_{i\in I}\neg(\exists\bar{y}_i\, \beta_i)
\end{array}\right]\Longrightarrow \neg\left[\begin{array}{l}
\exists\bar{x}'\bar{x}''\,\alpha'\wedge\alpha''\wedge\\[2mm]
\bigwedge_{i\in I}\neg(\exists\bar{x}'''\bar{y}_i\,
\alpha'''\wedge\beta_i)\end{array}\right]\] where the formula
$\exists\bar{x}\,\alpha$ is equivalent in $T$ to a decomposed
formula of the form
$\exists\bar{x}'\,\alpha'\wedge(\exists\bar{x}''\,\alpha''\wedge(\exists\bar{x}'''\,\alpha'''))$
with $\exists\bar{x}'\,\alpha'\in A'$,
$\exists\bar{x}''\,\alpha''\in A''$,
$\exists\bar{x}'''\,\alpha'''\in A'''$ and
$\exists\bar{x}'''\,\alpha'''$ different from
$\exists\varepsilon\,\vrai$.

Let us show the correctness of this rule. According to the conditions of application of this rule,  the formula
$\exists\bar{x}\,\alpha$ is equivalent in $T$ to a decomposed
formula of the form
$\exists\bar{x}'\,\alpha'\wedge(\exists\bar{x}''\,\alpha''\wedge(\exists\bar{x}'''\,\alpha'''))$
with $\exists\bar{x}'\,\alpha'\in A'$,
$\exists\bar{x}''\,\alpha''\in A''$,
$\exists\bar{x}'''\,\alpha'''\in A'''$ and
$\exists\bar{x}'''\,\alpha'''$ different from
$\exists\varepsilon\,\vrai$. Thus, the left formula of this rewriting rule is
equivalent in $T$ to the formula
\[\neg(\exists\bar{x}'\,\alpha'\wedge(\exists\bar{x}''\alpha''\wedge(\exists\bar{x}'''\alpha'''\wedge
\bigwedge_{i\in I}\neg(\exists\bar{y}_i\, \beta_i)))).\]  
Since $\exists\bar{x}'''\,\alpha'''\in A'''$,  then according to the fourth
point of Definition \ref{decomp} we have
$T\models\exists!\bar{x}'''\alpha'''$, thus using Corollary
\ref{unique1} the preceding formula is equivalent in $T$ to
\[\neg(\exists\bar{x}'\,\alpha'\wedge(\exists\bar{x}''\alpha''\wedge
\bigwedge_{i\in
I}\neg(\exists\bar{x}'''\alpha'''\wedge(\exists\bar{y}_i\,
\beta_i))))\] According to the definition of the working
formula the quantified variables have distinct names and different
from the names of the free variables, thus we can lift the
quantifications and then the preceding  formula is equivalent in
$T$ to
\[\neg(\exists\bar{x}'\,\alpha'\wedge(\exists\bar{x}''\alpha''\wedge
\bigwedge_{i\in
I}\neg(\exists\bar{x}'''\bar{y}_i\,\alpha'''\wedge\beta_i)))\]
i.e. to
\[\neg(\exists\bar{x}'\bar{x}''\,\alpha'\wedge\alpha''\wedge
\bigwedge_{i\in
I}\neg(\exists\bar{x}'''\bar{y}_i\,\alpha'''\wedge\beta_i)^*),\]
where the formula $(\exists\bar{x}'''\bar{y}_i\, \alpha'''\wedge\beta_i)^*$ is the formula $(\exists\bar{x}'''\bar{y}_i\, \alpha'''\wedge\beta_i)$ in which we have renamed the variables of $\bar{x}'''$ by distinct names and different from the names of the free variables. Thus, the rewriting rule (3) is correct in $T$.

\subsubsection*{Correctness of the rule (4):}

\[ \neg\left[\begin{array}{l}
\exists\bar{x}\,\alpha\wedge\\[2mm]
\bigwedge_{i\in I}\neg(\exists\bar{y}'_i\, \beta'_i)
\end{array}\right]\Longrightarrow \neg\left[\begin{array}{l}
\exists\bar{x}'\,\alpha'\wedge\\[2mm]
\bigwedge_{i\in
I'}\neg(\exists\bar{y}'_i\,\beta'_i)\end{array}\right]\] where the
formula $\exists\bar{x}\,\alpha$ is not an element of $A'$ and is
equivalent in $T$ to a decomposed formula of the form
$\exists\bar{x}'\,\alpha'\wedge(\exists\bar{x}''\,\alpha''\wedge(\exists\varepsilon\,\vrai))$
with $\exists\bar{x}'\,\alpha'\in A'$ and
$\exists\bar{x}''\,\alpha''\in A''$.  Each formula
$\exists\bar{y}'_i\,\beta'_i$ is an element of $A'$. $I'$ is the
set of the $i\in I$ such that $\exists\bar{y}'_i\beta'_i$ does not
have free occurrences of any variable of $\bar{x}''$.

Let us show the correctness of this rule. According to the conditions of application of this rule,  the formula
$\exists\bar{x}\,\alpha$ is equivalent in $T$ to a decomposed
formula of the form
$\exists\bar{x}'\,\alpha'\wedge(\exists\bar{x}''\,\alpha''\wedge(\exists\varepsilon\,\vrai))$
with $\exists\bar{x}'\,\alpha'\in A'$ and 
$\exists\bar{x}''\,\alpha''\in A''$. Moreover,  each formula
$\exists\bar{y}'_i\,\beta'_i$ belongs to $A'$.  Thus, the left formula of this rewriting rule is
equivalent in $T$ to the formula
\[\neg(\exists\bar{x}'\,\alpha'\wedge(\exists\bar{x}''\alpha''\wedge
\bigwedge_{i\in I}\neg(\exists\bar{y}'_i\, \beta'_i)))\] Let us
denote by $I_{1}$, the set of the $i \in I$ such that $x''_n$ does
not have free occurrences in the formula
$\exists\bar{y}'_i\beta'_i$, thus the preceding formula is
equivalent in $T$ to
\begin{equation}\label{kh1}
\neg(\exists\bar{x}'\alpha'\wedge(\exists x''_1...\exists
x''_{n-1}\left[\begin{array}{@{}l@{}}(\bigwedge_{i\in
I_1}\neg(\exists\bar{y}'_i\beta'_i))\wedge\\(\exists
x''_n\,\alpha''\wedge\bigwedge_{i\in
I-I_1}\neg(\exists\bar{y}'_i\beta'_i))\end{array}\right])).
\end{equation}
Since $\exists\bar{x}''\alpha''\in A''$ and
$\exists\bar{y}'_i\beta'_i\in A'$ for every $i\in I-I_1$, then according to Property
\ref{linfini} and the conditions 2 and 3 of Definition
\ref{decomp}, the formula (\ref{kh1}) is equivalent in $T$ to
\begin{equation}\label{kh2}
\textstyle{\neg(\exists\bar{x}'\alpha'\wedge(\exists
x''_1...\exists x''_{n-1}\,(\vrai\wedge\bigwedge_{i\in
I_1}\neg(\exists\bar{y}'_i\beta'_i)))).}
\end{equation}
By repeating the three preceding steps $(n-1)$ times, by denoting
by $I_{k}$ the set of the $i\in I_{k-1}$ such that $x''_{(n-k+1)}$
does not have free occurrences in $\exists\bar{y}'_i\beta'_i$, and
by using $(n-1)$ times Property \ref{zebb}, the preceding formula
is equivalent in $T$ to
\[
\textstyle{\neg(\exists\bar{x}'\alpha'\wedge\bigwedge_{i\in
I_n}\neg(\exists\bar{y}'_i\beta'_i)),}
\]
Thus, the rule (4) is correct in $T$.

\subsubsection*{Correctness of the rule (5):}
\[\neg\left[\begin{array}{l} \exists\bar{x}\,\alpha\wedge
\varphi\wedge\\[2mm] \neg\left[\begin{array}{l} \exists\bar{y}'\,
\beta'\wedge\\[2mm]\bigwedge_{i\in I}\neg(\exists\bar{z}'_i\,\delta'_i)
\end{array}\right]\end{array}\right]\Longrightarrow \left[\begin{array}{l}
\neg(\exists\bar{x}\,\alpha\wedge\varphi\wedge\neg(\exists\bar{y}'\,\beta'))\wedge\\[2mm]\bigwedge_{i\in I}\neg(\exists\bar{x}\bar{y}'\bar{z}'_i\,\alpha\wedge\beta'\wedge \delta'_i\wedge\varphi)^*\\
\end{array}\right]\]
where $I\neq\emptyset$ and the formulas $\exists\bar{y}'\,\beta'$
and $\exists\bar{z}'_i\,\delta'_i$ are elements of $A'$ for all $i\in I$.

Let us show the correctness of this rule. Since
$\exists\bar{y}'\beta'\in A'$ then according to the second point
of Definition \ref{decomp} we have
$T\models\exists?\bar{y}'\beta'$, thus using Corollary \ref{r2}
the preceding formula is equivalent in $T$ to
\[\neg\left[\begin{array}{l} \exists\bar{x}\,\alpha\wedge
\varphi\wedge\\[2mm] \neg\left[\begin{array}{l} (\exists\bar{y}'\,
\beta')\wedge\bigwedge_{i\in
I}\neg(\exists\bar{y}'\,\beta'\wedge(\exists\bar{z}'_i\,\delta'_i))
\end{array}\right]\end{array}\right]\]
According to the definition of  working formula the
quantified variables have distinct names and different from the
names of the free variables, thus we can lift the quantifications
and then the preceding  formula is equivalent in $T$ to
\[\neg\left[\begin{array}{l} \exists\bar{x}\,\alpha\wedge
\varphi\wedge\\[2mm] \neg\left[\begin{array}{l} (\exists\bar{y}'\,
\beta')\wedge\bigwedge_{i\in
I}\neg(\exists\bar{y}'\bar{z}'_i\,\beta'\wedge\delta'_i)
\end{array}\right]\end{array}\right]\]
thus to
\[\neg\left[\begin{array}{l} \exists\bar{x}\,\alpha\wedge
\varphi\wedge\\[2mm] \left[\begin{array}{l} (\neg(\exists\bar{y}'\,
\beta'))\vee\bigvee_{i\in
I}(\exists\bar{y}'\bar{z}'_i\,\beta'\wedge\delta'_i)
\end{array}\right]\end{array}\right]\]
After having distributed the $\wedge$ on the $\vee$ and lifted the
quantification $\exists\bar{y}'\bar{z}'_i$ we get
\[\neg\left[\begin{array}{l} (\exists\bar{x}\,\alpha\wedge
\varphi\wedge\neg(\exists\bar{y}'\, \beta'))\vee\\[2mm]\bigvee_{i\in
I}(\exists\bar{x}\bar{y}'\bar{z}'_i\,\alpha\wedge\varphi\wedge\beta'\wedge\delta'_i)
\end{array}\right]\]
which is equivalent in $T$ to \begin{equation}\label{encore}
\left[\begin{array}{l}
\neg(\exists\bar{x}\,\alpha\wedge\varphi\wedge\neg(\exists\bar{y}'\,\beta'))\wedge\\[2mm]\bigwedge_{i\in I}\neg(\exists\bar{x}\bar{y}'\bar{z}'_i\,\alpha\wedge\varphi\wedge\beta'\wedge \delta'_i)\\
\end{array}\right]\end{equation}
In order to satisfy the definition of the working formulas we
must rename the variables of $\bar{x}$ and $\bar{y}'$ by distinct
names and different from the names of the free variables. Let us
denote by
$(\exists\bar{x}\bar{y}'\bar{z}'_i\,\alpha\wedge\varphi\wedge\beta'\wedge
\delta'_i)^*$ the formula
$(\exists\bar{x}\bar{y}'\bar{z}'_i\,\alpha\wedge\varphi\wedge\beta'\wedge
\delta'_i)$ in which we have renamed the variables of $\bar{x}$
and $\bar{y}'$ by distinct names and different from the names of the free variables. Thus, the formula (\ref{encore}) is equivalent in
${ T}$ to \[\left[\begin{array}{l}
\neg(\exists\bar{x}\,\alpha\wedge\varphi\wedge\neg(\exists\bar{y}'\,\beta'))\wedge\\[2mm]\bigwedge_{i\in I}\neg(\exists\bar{x}\bar{y}'\bar{z}'_i\,\alpha\wedge\varphi\wedge\beta'\wedge \delta'_i)^*\\
\end{array}\right]\]
 Thus, the
rule (5) is correct in $T$. 
$\\[3mm]${\emph{Proof, third part:}} Every finite 
application of the rewriting rules on a working formula 
produces a wnfv conjunction of solved formulas.

Recall that  we
write $\bigwedge_{i\in I}\varphi_i$, and call \emph{conjunction}
each formula of the form
$\varphi_{i_1}\wedge\varphi_{i_2}\wedge...\wedge\varphi_{i_n}\wedge\vrai$.
In particular, for $I=\emptyset$, the conjunction $\bigwedge_{i\in
I}\varphi_i$ is reduced to $\vrai$. Moreover, we do not
distinguish two formulas which can be made equal using the
following transformations of  sub-formulas:
\[\begin{array}{@{}c@{}}\varphi\wedge\varphi\Longrightarrow\varphi,\;\;\varphi\wedge\psi\Longrightarrow\psi\wedge\varphi,\;\;(\varphi\wedge\psi)\wedge\phi\Longrightarrow\varphi\wedge(\psi\wedge\phi),\\\varphi\wedge\vrai\Longrightarrow\varphi,\;\;\varphi\vee\faux\Longrightarrow\varphi.\end{array}\]

Let us show first that every substitution of a sub-working formula of a conjunction of working formulas by a conjunction of working formulas produces a conjunction of working formulas.  Let $\bigwedge_{i\in I}\varphi_i$ be a conjunction of working formulas. Let $\varphi_k$ with $k\in I$ be an element of this conjunction  of depth $d_k$. Two cases arise: \begin{enumerate} \item 
 We replace $\varphi_k$ by a conjunction of working formulas. Thus, let $\bigwedge_{j\in J_k}\phi_j$ be a conjunction of working formulas which is equivalent to $\varphi_k$ in $T$. The conjunction of working formulas $\bigwedge_{i\in I}\varphi_i$ is equivalent in $T$ to \[(\bigwedge_{i\in I-\{k\}}\varphi_i)\wedge(\bigwedge_{j\in J_k}\phi_j)\] which is clearly a conjunction of working formulas.
\item
 We replace a strict sub-working formula of $\varphi_k$ by a conjunction of working formulas. Thus, let $\phi$ be a sub-working formula of $\varphi_k$ of depth $d_{\phi}<d_k$ (thus $\phi$ is different from $\varphi_k$). Thus, $\varphi_k$ has a sub-working formula\footnote{By considering that the set of the sub-formulas of any formula $\varphi$ contains also the whole formula $\varphi$.} of the form 
\[\neg(\exists\bar{x}\alpha\wedge(\bigwedge_{l\in L} \psi_l)\wedge\phi),
\]
where $L$ is a finite (possibly empty) set and all the $\psi_l$ are working formulas. Let $\bigwedge_{j\in J}\phi_j$ be a conjunction of working formulas which is equivalent to $\phi$ in $T$.  Thus the preceding sub-working formula of $\varphi_k$ is equivalent in $T$ to 
\[\neg(\exists\bar{x}\alpha\wedge(\bigwedge_{l\in L} \psi_l)\wedge(\bigwedge_{j\in J}\phi_j)),
\]
which is clearly a sub-working formula and thus $\varphi_k$ is equivalent to a working formula and thus  $\bigwedge_{i\in I}\varphi_i$ is equivalent to a conjunction of working formulas.
\end{enumerate}
 From 1 and 2 we deduce that  (i) every substitution of a sub-working formula of a conjunction of working formulas by a conjunction of working formulas produces a conjunction of working formulas. 

 Since each rule transforms a working formula into a conjunction of working formulas, then according to (i)  every finite application of the rewriting rules on a working formula produces a conjunction of working formulas. Let us show now that each of these final working formulas is solved. 

Let $\varphi$ be a working formula. According to all what we have shown, every finite application of our rules on $\varphi$ produces a conjunction $\phi$ of working formulas. Suppose that the rules terminate  and one of the working formulas of $\phi$ is not solved.  Let $\psi$ be this formula, two cases arise:

{\bfseries Case 1:} $\psi$ is a working formula of depth
greater than 2. Thus, $\psi$ has a sub-formula of the form
\[\neg\left[\begin{array}{l} \exists\bar{x}\,\alpha\wedge
\psi_1\wedge\\[2mm] \neg\left[\begin{array}{l} \exists\bar{y}\,
\beta\wedge\bigwedge_{i\in I}\neg(\exists\bar{z}_i\,\delta_i)
\end{array}\right]\end{array}\right]\]
where $\psi_1$ is a conjunction of working formulas, $I$ is a
nonempty set  and $\alpha$, $\beta$ and $\delta_i$ are elements of $A$ for all $i\in I$.
Let
$(\exists\bar{y}'\beta'\wedge(\exists\bar{x}''\beta''\wedge(\exists\bar{y}'''\beta''')))$
be the decomposed formula in $T$ of $\exists\bar{y}\beta$ and let
$(\exists\bar{z}'_i\delta'_i\wedge(\exists\bar{z}''_i\delta''_i\wedge(\exists\bar{z}'''_i\delta'''_i)))$
be the decomposed formula in $T$ of $\exists\bar{z}_i\delta_i$. If
$\exists\bar{y}'''\beta'''$ is not the formula
$\exists\varepsilon\vrai$ then the rule (3) can still be applied
which contradicts our supposition. Thus, suppose that
\begin{equation}\label{zb1}
\exists\bar{y}'''\beta'''=\exists\varepsilon\vrai\end{equation} If
there exists $k\in I$ such that $\exists\bar{z}'''_k\delta'''_k$
is not the formula $\exists\varepsilon\vrai$ then the rule (3) can
be still applied (with $I=\emptyset$) which contradicts our
supposition. Thus, suppose that
\begin{equation}\label{zb2}
\exists\bar{z}'''_i\delta'''_i=\exists\varepsilon\vrai\end{equation}
for all $i\in I$. If there exists $k\in I$ such that $\exists\bar{z}_k\delta_k$ is
not an element of $A'$ then since we have (\ref{zb2}), the rule
(4) can still be applied (with $I=\emptyset$) which contradicts
our supposition. Thus, suppose that
\begin{equation}\label{zb3} \exists\bar{z}_i\delta_i\in A'\end{equation}for all $i\in I$. If $\exists\bar{y}\beta$ is
not an element of $A'$ then since we have (\ref{zb1}) and
(\ref{zb3}), the rule (4) can still be applied  which contradicts
our supposition. Thus, suppose that
\begin{equation}\label{zb4}
\exists\bar{y}\beta\in A'
\end{equation}
Since we have (\ref{zb3}) and (\ref{zb4}) then the rule (5) can
still be applied  which contradicts all our suppositions.

{\bfseries Case 2:} $\psi$ is a working formula of the form
\[\neg(\exists\bar{x}\,
\alpha\wedge\bigwedge_{i\in I}\neg(\exists\bar{y}_i\,\beta_i))\]
where at least one of the following conditions holds:
\begin{enumerate}
\item $\alpha$ is the formula $\faux$, \item there exists $k \in
I$ such that $\beta_k$ is the formula $\vrai$ or $\faux$, \item
there exists $k \in I$ such that $\exists\bar{y}_k\beta_k\not\in
A'$, \item $\exists\bar{x}\alpha\not\in A'$. \end{enumerate} If
the condition (1) holds then the rule (2) can still be applied
which contradicts our suppositions. If the condition (2) holds
then the rules (1) and (2) can still be applied which contradicts
our suppositions. If the condition (3) holds then the rule (3) or
(4) (with $I=\emptyset$) can still be applied which contradicts
our suppositions. If the condition
(4) holds then according to the preceding point
$\exists\bar{y}_i\beta_i\in A'$ for all $i\in I$ and thus the rule (3) or (4) can
still be applied which contradicts our suppositions.

From Case 1 and Case 2, our suppositions are always false thus
$\psi$ is a solved formula and thus $\phi$ is a conjunction of
solved formulas.

\subsection{The algorithm of resolution}
Having any formula $\psi$, the resolution of $\psi$ proceeds as
follows:
\begin{enumerate}
\item Transform the formula $\psi$ into a normalized formula and then into a working formula
$\varphi$ which is wnfv and equivalent to $\psi$ in $T$. \item Apply the preceding
rewriting rules on $\varphi$ as many time as possible. At the end
we obtain a conjunction $\phi$ of solved formulas.
\end{enumerate}
According to Property \ref{gros}, the application of the rewriting
rules on a formula $\psi$ without free variables produces a
conjunction $\phi$ of solved formulas
which is equivalent to $\psi$ in $T$ and does not contain free variables. According to Property \ref{uniq},
$\phi$ is either the formula $\vrai$ or $\neg\vrai$, thus
either $T\models\psi$ or $T\models\neg\psi$ and thus $ T$ is a
complete theory. We can now present our main result:
\begin{corollaire}\label{thm}
If $T$ is a decomposable theory then every formula is equivalent
in $T$ either to $\vrai$ or to $\faux$ or to a Boolean combination
of elements of $A'$ which has at least one free variable.
\end{corollaire}

\begin{remarque}
There exists another way to solve the first-order formulas in $T$
specially in the case where there exists at least one free variable
in the initial formula $\psi$ and when the goal of the resolution
is to have explicit and understanding solutions of these free
variables in $\psi$. In this case it is better to run the
preceding  algorithm on $\neg\psi$. Let then \[\bigwedge_{i\in
I}\neg(\exists\bar{x}'_i\,\alpha'_i\wedge\bigwedge_{j\in
J_i}\neg(\exists\bar{y}'_{ij}\,\beta'_{ij}))\] be the conjunction
of solved formulas obtained by application of the preceding  rules
on $\neg\psi$. The formula
\[\bigvee_{i\in
I}(\exists\bar{x}'_i\,\alpha'_i\wedge\bigwedge_{j\in
J_i}\neg(\exists\bar{y}'_{ij}\,\beta'_{ij}))\] is a wnfv
disjunction of formulas which is equivalent to $\psi$ in $T$. It is more easy to understand the solutions of the free variables of this disjunction of  solved formulas  than those of  a conjunction of solved formulas.
\end{remarque}

\section{The theory ${\cal T}$ of finite or infinite trees}\label{arbre}

\subsection{The axioms}
The theory ${\cal T}$ of finite or  infinite  trees built on an
{\bfseries infinite} set $\Ff$ of distinct function symbols has as axioms the infinite
set of propositions of one of the three following forms:

\[\begin{tabular}{lll}
$\forall\bar{x}\forall\bar{y}\hspace{5mm} $& $\neg f\bar{x}=g\bar{y} $&\hspace{10mm} [1] \\
$\forall\bar{x}\forall\bar{y} $\hspace{5mm}& $f\bar{x}=f\bar{y}\rightarrow \bigwedge_i x_i=y_i$  &\hspace{10mm} [2] \\
$\forall\bar{x}\exists!\bar{z}$\hspace{5mm}& $\bigwedge_i z_i=t_i[\bar{x}\bar{z}]$ & \hspace{11mm}[3] \\
\end{tabular}
\]
where $f$ and $g$ are distinct function symbols taken from $\Ff$,
$\bar{x}$ is a vector of possibly non-distinct  variables $x_i$, $\bar{y}$ is a vector of possibly non-distinct  variables $y_i$, $\bar{z}$ is a
vector of distinct variables $z_i$ and $t_i[\bar{x}\bar{z}]$ is a
term which begins with an element of $\Ff$ followed by variables
taken from $\bar{x}$ or $\bar{z}$.  Note that this theory does not
accept full elimination of quantifiers. In fact, in the formula $\exists
x\,y=f(x)$ we can not remove or eliminate the quantifier $\exists
x$.

\subsection{Properties of
${\cal T}$} Suppose that the variables of $V$ are ordered by a strict linear dense order relation without endpoints denoted by
$\succ$.
\begin{definition1}\label{flat} A conjunction $\alpha$ of flat equations
is called \emph{$(\succ\!)$-solved} if all its left-hand sides are
distinct and $\alpha$ does not contain equations of the form $x=x$
or $y = x$, where $x$ and $y$ are variables such that $x\succ y$.
\end{definition1}
\begin{propriete}\label{resolution}
Every conjunction $\alpha$ of flat formulas is equivalent in
${\cal T}$ either to false or to a $(\succ\!)$-solved conjunction
of flat equations.
\end{propriete}

\begin{proof}
To prove this property we introduce the following rewriting rules:

\[\begin{array}{llll}
(1)\;\;\;& \faux\wedge\alpha  &\;\;\;\Longrightarrow \;\;\;& \faux,\\[1mm]
(2)& x=fy_1...y_m \wedge x=gz_1...z_n &\;\;\;\Longrightarrow & \faux,\\[1mm]
(3)& x=fy_1...y_n\wedge x=fz_1...z_n & \;\;\;\Longrightarrow & x=fy_1...y_n\wedge\bigwedge_{i\in \{1,...,n\}}y_i=z_i,\\[1mm]
(4)& x=x  &\;\;\;\Longrightarrow &\vrai \\[1mm] (5)& y=x
&\;\;\;\Longrightarrow & x=y\\[1mm]
(6)& x=y\wedge x=fz_1...z_n &\;\;\;\Longrightarrow & x=y\wedge
y=fz_1...z_n\\[1mm]
(7) &x=y \wedge x=z &\;\;\;\Longrightarrow & x=y\wedge y=z
\end{array}\]
with $\alpha$ any formula and $f$ and $g$ two distinct function
symbols taken from $\Ff$. The rules (5), (6) and (7) are applied
only if $x\succ y$. This condition prevents infinite loops.

 Let us prove now that
every repeated application of the preceding rewriting rules on any
conjunction $\alpha$ of flat formulas, is terminating and
producing either the formula $\faux$ or a $(\succ\!)$-solved
conjunction of flat equations which is equivalent to $\alpha$ in ${\cal
T}$. $\\$ {\emph{Proof, first part:}} The application of the
rewriting rules terminates. Since the variables which occur in our
formulas are ordered by the strict linear  order relation without endpoints
$``\succ"$, we can number them by positive integers such that
\[x\succ y \leftrightarrow no(x)>no(y),\] where $no(x)$ is the
number associated  to the variable $x$. Let us consider the
4-tuple $(n_1,n_2,n_3,n_4)$ where the $n_i$'s are the following
positive integers:
\begin{itemize}
\item $n_1$ is the number of occurrences of sub-formulas of the
form $x=fy_1...y_n$, with $f\in\Ff$, \item $n_2$ is the number of
occurrences of atomic formulas, \item $n_3$ is the sum of the
$no(x)$'s for all occurrences of a variable $x$, \item $n_4$ is the
number of occurrences of formulas of the form $y=x$, with $x\succ
y$.
\end{itemize}
For each rule, there exists an integer  $i$ such that the application of
this rule decreases or does not change the values of the $n_j$'s,
with $1\leq j<i$, and decreases the value of $n_i$. This integer $i$ is
equal to:  2 for the rule (1), 1 for the rules (2) and (3), 3 for
the rules (4), (6) and (7), 4 for the rule (5). To each sequence
of formulas obtained by a finite application of the preceding
rewriting rules, we can associate a series of 4-tuples
$(n_1,n_2,n_3,n_4)$ which is strictly decreasing in the
lexicographic order. Since the $n_i$'s are positive integers, they
cannot be negative, thus this series of 4-tuples is a finite
series and the application of the rewriting rules terminates.
$\\[3mm]${\emph{Proof, second part:}} The rules preserve equivalence in
${\cal T}$. The rule (1) is evident in ${\cal T}$. The rules (2)
preserves the equivalence in ${\cal T}$ according to the axiom 1.
The rule (3) preserves the equivalence in ${\cal T}$ according to
the axiom 2. The
rules (4), (5), (6) and (7) are evident in ${\cal T}$. $\\[3mm]${\emph{Proof, third part:}} The
application of the rewriting rules terminates either by $\faux$ or
by a $(\succ\!)$-solved conjunction of flat equations. Suppose
that the application of the rewriting rules on a conjunction
$\alpha$ of flat formulas terminates by a formula $\beta$ and at
least one of the following conditions holds:
\begin{enumerate}\item $\beta$ is not the formula $\faux$ and has at least a sub-formula of the form $\faux$,
\item $\beta$ has two equations with the same left-hand side, \item
$\beta$ contains equations of the form $x=x$ or $y=x$ with $x\succ
y$. \end{enumerate} If the condition 1 holds then the rule (1) can
still be applied which contradicts our supposition. If the
condition 2 holds then the rules (2), (3), (6) and (7) can still
be applied which contradicts our supposition. If the condition 3
holds then the rules (4) and (5) can still be applied which
contradicts our supposition. Thus, the formula $\beta$ according to
Definition \ref{flat} is either the formula $\faux$ or a
$(\succ\!)$-solved conjunction of flat equations.
\end{proof}

Let us introduce now the notion of \emph{reachable variable} and
\emph{reachable equation}.
\begin{definition1}\label{accesi}
The equations and variables reachable from the variable $u$ in the
formula \[\exists\bar{x}\,\bigwedge_{i=1}^{n} v_i=t_i\] are those
who occur in at least one of its sub-formulas of the form
$\bigwedge_{j=1}^{m} v_{k_j}=t_{k_j}$, where $v_{k_1}$ is the
variable $u$ and $v_{k_j+1}$ occurs in the term $t_{k_j}$ for all
$j\in\{1,..,m\}$. The equations and variables reachable of this
formula are those who are reachable from a variables which does
not occur in $\bar{x}$.
\end{definition1}
\begin{example}
In the formula  \[\exists uvw\, z = fuv \wedge v = gvu \wedge w =
fuv,\] the equations $z = fuv$ and $v = gvu$ and the variables $u$
and $v$ are reachable. On the other hand the equation $w = fuv$
and the variable $w$ are not reachable.
\end{example}
 According to the axioms [1] and [2] of ${\cal T}$ we have the following property
\begin{propriete}\label{acc}
Let $\alpha$ be a conjunction of flat equations. If all the
variables of $\bar{x}$ are reachable in $\exists\bar{x}\,\alpha$
then ${\cal T}\models\exists?\bar{x}\,\alpha$.
\end{propriete} According to the axiom 3 we have:
\begin{propriete}\label{unique} Let  $\alpha$ be a $(\succ\!)$-solved conjunction of flat equations and let $\bar{x}$ be the vector of its left-hand sides.
We have ${\cal T}\models\exists!\bar{x}\,\alpha$.
\end{propriete}

\subsection{${\cal T}$ is decomposable}

\begin{propriete}\label{th}
${\cal T}$ is a decomposable theory. \end{propriete} Let us show
that ${\cal T}$ satisfies the conditions of Definition
\ref{decomp}.

\subsubsection{Choice of the sets $\Psi(u)$, $A$, $A'$, $A''$ and
$A'''$} Let $F_0$ be the set of the $0$-ary function symbols of $F$. The sets $\Psi(u)$, $A$, $A'$, $A''$ and
$A'''$ are chosen as follows:
 \begin{itemize}
 \item
 $\Psi(u)$ is the set $\{\faux\}$ if $F-F_0=\emptyset$, else it contains  formulas of the form
$\exists\bar{y}\,u=f\bar{y}$ with $f\in \Ff-F_0$,\item $A$ is the set $FL$, \item $A'$ is the
set of  formulas of the form $\exists\bar{x}'\alpha'$ such that
\begin{itemize}\item $\alpha'$ is either the formula $\faux$ or a $(\succ\!)$-solved conjunction
of flat equations where the order $\succ$ is such that all the
variables of $\bar{x}'$ are greater than the free variables of
$\exists\bar{x}'\alpha'$, \item all the variables of $\bar{x}'$
and all the equations of $\alpha'$ are reachable in
$\exists\bar{x}'\alpha'$, \end{itemize}\item
 $A''$ is the set of  formulas of the form $\exists\bar{x}''\,\vrai$, \item
 $A'''$ is the
set of formulas of the form $\exists\bar{x}'''\alpha'''$ such
that $\alpha'''$ is a $(\succ\!)$-solved conjunction of flat
equations and $\bar{x}'''$ is the vector of the left-hand sides of
the equations of $\alpha'''$.
\end{itemize}

It is clear that $FL$ is ${\cal T}$-closed and $A'$, $A''$ and $A'''$ contain formulas of the form $\exists\bar{x}\,\alpha$ with $\alpha\in FL$. Let us now show that ${\cal T}$ satisfies the five condition of
Definition \ref{decomp} \subsubsection{\boldmath ${\cal
T}$\label{dex} satisfies the first condition} Let us show that
every formula of the form $\exists\bar{x}\,\alpha\wedge\psi$, with
$\alpha\in FL$ and $\psi$ any formula, is equivalent in ${\cal T}$
to a wnfv formula of the form
\begin{equation}\label{an1980}
\exists \bar{x}'\,\alpha'\wedge(
\exists\bar{x}''\,\alpha''\wedge(\exists\bar{x}'''\,\alpha'''\wedge
\psi)),
\end{equation}
with $\exists\bar{x}'\,\alpha'\in A'$,
$\exists\bar{x}''\,\alpha''\in A''$ and
$\exists\bar{x}'''\,\alpha'''\in A'''$.

Let us choose the order $\succ$ such that all the variables of
$\bar{x}$ are greater than the free variables of
$\exists\bar{x}\alpha$.  According to Property
\ref{resolution} two cases arise:

Either $\alpha$ is equivalent to $\faux$ in ${\cal T}$. Thus,  $\bar{x}'=\bar{x}''=\bar{x}'''=\varepsilon$, $\alpha'=\faux$ and
$\alpha''=\alpha'''=\vrai$.

Or, $\alpha$ is equivalent to a $(\succ\!)$-solved
conjunction $\beta$ of flat equations. Let $X$ be the set of the variables
 of the vector $\bar{x}$. Let  $Y_{rea}$ be the set of the reachable variables of
 $\exists\bar{x}\beta$. Let $Lhs$ be the set of the variables which occur in a left-hand side of an equation of
 $\beta$. We have: $\\$
$\:\:-\:$ $\bar{x}'$ contains the variables of $X\cap Y_{rea}$.\\
$\:\:-\:$ $\bar{x}''$ contains the variables of $(X-Y_{rea})-Lhs$.\\
$\:\:-\:$ $\bar{x}'''$ contains the variables of $(X-Y_ {rea})\cap Lhs$.\\
$\:\:-\:$ $\alpha'$ is the conjunction of the reachable
equations of $\exists\bar{x}\beta$.\\
$\:\:-\:$ $\alpha''$ is the formula $\vrai$.\\
 $\:\:-\:$ $\alpha'''$ is the conjunction of the unreachable
equations of $\exists\bar{x}\beta$.\\

According to our construction it is clear that  $\exists\bar{x}'\alpha'\in A'$, $\exists\bar{x}''\alpha''\in A''$ and  $\exists\bar{x}'''\alpha\in A'''$. Let us show that (\ref{an1980}) and $\exists\bar{x}\alpha\wedge\psi$ are equivalent in ${\cal T}$. Let ${X}'$, ${X}''$ and ${X}'''$ be the sets of the variables of the vectors  $\bar{x}'$, $\bar{x}''$ and  $\bar{x}'''$. If $\alpha$ is equivalent to $\faux$ in ${\cal T}$ then the equivalence of the decomposition is evident. Else  $\beta$ is a conjunction of flat equations and thus according to our construction we have:      ${X}={X}'\cup{X}''\cup{X}'''$, ${X}'\cap{X}''=\emptyset$, ${X}'\cap{X}'''=\emptyset$, ${X}''\cap{X}'''=\emptyset$, for all $x''_i\in {X}''$ we have  $x''_i\not\in var(\alpha')$ and for all $x'''_i\in{X}'''$ we have  $x'''_i\not\in var(\alpha'\wedge\alpha'')$. Moreover each equation of $\beta$ occurs  in $\alpha'\wedge\alpha''\wedge\alpha'''$ and each equation in $\alpha'\wedge\alpha''\wedge\alpha'''$ occurs in $\beta$ and thus ${\cal T}\models\beta\leftrightarrow(\alpha'\wedge\alpha''\wedge\alpha''')$. We have shown that the vectorial quantifications  are coherent and the equivalence ${\cal T}\models\beta\leftrightarrow\alpha'\wedge\alpha''\wedge\alpha'''$ holds. According to Property \ref{resolution} we have ${\cal T}\models\alpha\leftrightarrow\beta$ and thus, the decomposition keeps the equivalence in ${\cal T}$.

\begin{example}
Let us decompose the following formula $\varphi$ \[\exists xyv\,
z=fxy\wedge z=fxw\wedge v=fz.\] First, since $w$ and $z$ are free
in $\varphi$ then the order $\succ$ will be chosen as follows:
\[x\succ y\succ v\succ w\succ z.\] Note that the quantified variables are greater than the free variables. Then, using the rewriting rules of
Property \ref{resolution} we transform the conjunction of
equations to a $(\succ\!)$-solved formula. Thus, the formula
$\varphi$ is equivalent in ${\cal T}$ to the following formula
$\psi$
\[\exists xyv\, z=fxy\wedge y=w\wedge v=fz.\]
Since the variables $x,y,w$ and the equations $z=fxy, y=w$ are
reachable in $\psi$ then $\psi$ is equivalent in ${\cal T}$ to the
following decomposed formula \[\exists xy\, z=fxy\wedge
y=w\wedge(\exists\varepsilon\,\vrai\wedge(\exists v\, v=fz)).\] It
is clear that $(\exists xy\,z=fxy\wedge y=w)\in A'$,
$(\exists\varepsilon\,\vrai)\in A''$ and $(\exists v\,v=fz)\in
A'''$.
\end{example}

\subsubsection{\boldmath ${\cal T}$ satisfies the second
condition}

Let us show that if $\exists\bar{x}'\alpha'\in A'$ then ${{\cal
T}}\models\exists?\bar{x}'\alpha'$. Since
$\exists\bar{x}'\alpha'\in A'$ and according to the choice of the
set $A'$, either $\alpha'$ is the formula $\faux$ and thus we have
immediately  ${\cal T}\models\exists?\bar{x}'\alpha'$ or $\alpha'$
is a $(\succ\!)$-solved conjunction of flat equations and the
variables of $\bar{x}'$ are reachable in $\exists\bar{x}'\alpha'$.
Thus, using Property \ref{acc} we get ${\cal
T}\models\exists?\bar{x}'\alpha'$.

Let us show now that if $y$ is a free variable of
$\exists\bar{x}'\alpha'$ then ${\cal
T}\models\exists?y\bar{x}'\,\alpha'$ or there exists
$\psi(u)\in\Psi(u)$ such that ${\cal T}\models\forall
y\,(\exists\bar{x}'\,\alpha')\rightarrow\psi(y)$. Let $y$ be a
free variable of $\exists\bar{x}'\alpha'$. It is clear that
$\alpha'$ can not be in this case the formula $\faux$. Thus, four
cases arise:

If $y$ occurs in a sub-formula of $\alpha'$ of the form
$y=t(\bar{x}',\bar{z}',y)$, where $\bar{z}'$ is the set of the
free variables of $\exists\bar{x}'\alpha'$ which are different
from $y$ and where $t(\bar{x}',\bar{z}',y)$ is a term which begins
by an element of $\Ff-F_0$ followed by variables taken from $\bar{x}'$
or $\bar{z}'$ or $\{y\}$, then the formula
$\exists\bar{x}'\alpha'$ implies in ${\cal T}$ the formula
$\exists\bar{x}'\, y=t(\bar{x}',\bar{z}',y),$ which implies in
${\cal T}$ the formula
$\exists\bar{x}'\bar{z}'w\,y=t(\bar{x}',\bar{z}',w),$
 where $y=t(\bar{x}',\bar{z}',w)$ is the formula $y=t(\bar{x}',\bar{z}',y)$
 in which we have replaced every free occurrence of $y$ in the term $t(\bar{x}',\bar{z}',y)$ by the variable $w$.
 According to the choice of the set $\Psi(u)$, the formula $\exists\bar{x}'\bar{z}' w\,u=t(\bar{x}',\bar{z}',w)$ belongs to
 $\Psi(u)$.

If $y$ occurs in a sub-formula of $\alpha'$ of the form $y=f_0$ with $f_0\in F_0$ then according to the third axiom of ${\cal T}$ we have ${\cal T}\models\exists!y\, y=f_0$. Thus (i) ${\cal T}\models\exists?y\, \alpha'$. On the other hand, since $\alpha'$ is
$(\succ\!)$-solved, $y$ has no occurrences in an other left-hand
side of an equation of $\alpha'$, thus since the variables of
$\bar{x}$ are reachable in $\exists\bar{x}'\alpha'$ (according to
the choice of the set $A'$), all the variables of $\bar{x}'$ keep
reachable in $\exists\bar{x}'y\,\alpha'$ and thus using (i) and Property \ref{acc} we get ${\cal T}\models\exists?\bar{x}'y\, \alpha'$.

If $y$ occurs in a sub-formula of $\alpha'$ of the form $y=z$
then: \begin{enumerate}\item According to the choice of the set
$A'$, the order $\succ$ is such that all the variables of
$\bar{x}'$ are greater than the free variables of
$\exists\bar{x}'\alpha'$. \item According to  Definition
\ref{resolution} of the $(\succ\!)$-solved formula, we have $y
\succ z$.\end{enumerate} From (1) and (2), we deduce that $z$ is a
free variable in $\exists\bar{x}'\alpha'$.  Since $\alpha'$ is
$(\succ\!)$-solved, $y$ has no occurrences in an other left-hand
side of an equation of $\alpha'$, thus since the variables of
$\bar{x}$ are reachable in $\exists\bar{x}'\alpha'$ (according to
the choice of the set $A'$), all the variables of $\bar{x}'$ keep
reachable in $\exists\bar{x}'y\,\alpha'$. More over, for each
value of $z$ there exists at most a value for $y$. Thus, using
Property \ref{acc} we get
 ${{\cal T}}\models\exists?\bar{x}'y\,\alpha'$.

If $y$ occurs only in the right-hand sides of the equations of
$\alpha'$ then according to the choice of the set $A'$, all the
variables of $\bar{x}'$ and all the equations of $\alpha'$ are
reachable in $\exists\bar{x}'\alpha'$. Thus, since $y$ does not
occur in a left-hand side of an equation of $\alpha'$, the
variable $y$ and the variables of $\bar{x}'$ are reachable in
$\exists\bar{x}'y\,\alpha'$ and thus using Property \ref{acc} we
get ${{\cal T}}\models\exists?\bar{x}'y\,\alpha'$. In all cases
${\cal T}$ satisfies the second condition of  Definition
\ref{decomp}.

\subsubsection{\boldmath ${\cal T}$ satisfies the third condition}

First, we present a property which hold in any model $M$ of ${\cal
T}$. This property results from the axiomatization of ${\cal T}$
(more exactly from axioms 1 and 2) and the infinite set $\Ff$ of function
symbols.

\begin{propriete}\label{dom}
Let $M$ be a  model of ${\cal T}$ and let $f$ be a function symbol
taken from $\Ff-F_0$. The set of the individuals $i$ of $M$, such that
$M\models \exists \overline{x}\,i=f\overline{x}$, is infinite.
\end{propriete}

Let $\exists\bar{x}''\alpha''$ be a formula which belongs to
$A''$. According to the choice of $A''$, this formula is
of the form $\exists\bar{x}''\,\vrai$. Let us show that, for every
variable $x''_j$ of $\bar{x}''$ we have ${{\cal
T}}\models\exists^{\Psi(u)}_{\infty}x_j\,\vrai$. Two cases arise:

If $F-F_0=\emptyset$ then $\Psi(u)=\{\faux\}$ and $F_0$ is infinite since the theory is defined on an infinite set of function symbols. According to axiom 1 of ${\cal T}$, for all distinct constants $f$ and $g$  correspond two distinct individuals in all models of ${\cal T}$. Thus, since $F_0$ is infinite there exists an infinite set of individuals in all models of ${\cal T}$ and thus according to Definition \ref{infini} we have: ${{\cal
T}}\models\exists^{\{\faux\}}_{\infty}x_j\,\vrai$. 

If $F-F_0\neq\emptyset$ then $\Psi(u)$ contains formulas of the form $\exists\bar{z}\,u=f\bar{z}$ with $f\in F-F_0$. Let $M$ be a
model of ${\cal T}$. Since the formula $\exists x''_j\,\vrai$ does
not have free variables, it is already instantiated, and thus
according to Definition \ref{infini} it is enough to show that
there exists an infinity of individuals $i$ of $M$ which satisfy
the following condition:
\begin{equation}\label{quatre}M\models \neg\psi_1(i)\wedge \cdots \wedge
\neg\psi_n(i),\end{equation} with $\psi_j(u)\in\Psi(u)$, i.e. of the form $\exists\bar{z}\,u=f\bar{z}$ with $f\in F-F_0$. Two cases arise:  
\begin{itemize}\item
 If $F-F_0$ is a finite set then $F_0$ is infinite because the  theory is defined on infinite set of function symbols. Thus, there exists an infinity of constants $f_k$ which are different from all the function symbols of all the $\psi_j(u)$ of (\ref{quatre}) and thus using  axiom 1 of ${\cal T}$  there exists an infinity of distinct individuals $i$ such that  (\ref{quatre}).
\item
If  $F-F_0$ is infinite then there exists a  formula $\psi(u)^*\in \Psi(u)$ which is different from all the $\psi_j(u)$ of (\ref{quatre}), i.e. which has a function symbol which is different from  the function symbols of all the $\psi_1(u)\cdots
\psi_n(u)$. According  to Property \ref{dom} there exists an infinity of individuals $i$ such that $M\models\psi(i)^*$. Since this $\psi(u)^*$ is different from all the $\psi_j(u)$, then according to axiom 1 of ${\cal T}$ there exists an infinite set of individuals $i$ such that $M\models\psi(i)^*\wedge \neg\psi_1(i)\wedge \cdots \wedge
\neg\psi_n(i)$ and thus such that  (\ref{quatre}).
\end{itemize}

\subsubsection{\boldmath ${\cal T}$ satisfies the fourth
condition} Let us show that if $\exists\bar{x}'''\alpha'''\in
A'''$ then ${\cal T}\models\exists!\bar{x}'''\,\alpha'''$.  Let
$\exists\bar{x}'''\alpha'''$ be an element of $A'''$. According to
the choice of the set $A'''$ and Property \ref{unique} we get
immediately ${\cal T}\models\exists!\bar{x}'''\alpha'''$.
\subsubsection{\boldmath ${\cal T}$ satisfies the fifth condition}
Let us show that if the formula $\exists\bar{x}'\alpha'$ belongs
to $A'$ and has no free variables then this formula is either the
formula $\exists\varepsilon\vrai$ or $\exists\varepsilon\faux$.
Let $\exists\bar{x}'\alpha'$ be a formula, without free variables,
which belongs to $A'$. We have \begin{enumerate}\item According to
the choice of the set $A'$, all the variables and equations of
$\exists\bar{x}'\alpha'$ are reachable in $\exists\bar{x}'\alpha'$
and $\alpha'$ is either the formula $\faux$ or a
$(\succ\!)$-solved conjunction of flat equations. \item Since the
formula $\exists\bar{x}'\alpha'$ has no free variables and
according to Definition \ref{accesi} there exists in this case
neither variables nor equations reachable in
$\exists\bar{x}'\alpha'$,\end{enumerate}Thus, From (1) and (2),
$\bar{x}'$ is the empty vector, i.e. $\varepsilon$ and $\alpha'$ is
either the formula $\vrai$ or $\faux$.

\subsection{Solving first-order formulas in ${\cal T}$}

 Since ${\cal T}$ is decomposable we can apply our general algorithm and solve any
first-order formula. Let us first recall the related works about
the resolution of tree constraints: the unification of finite
terms, i.e. the resolution of conjunctions of equations in the
theory of finite trees has first been studied by A.
Robinson~\cite{rob43}. Some better algorithms with better
complexities has been proposed after by M.S. Paterson and
M.N.Wegman~\cite{pat41} and A. Martelli and U.
Montanari~\cite{mat39}. The resolution of conjunctions of
equations in the theory of infinite trees has been studied by G.
Huet~\cite{hue}, by A. Colmerauer~\cite{col7,col84} and by J.
Jaffar~\cite{jaf28}. The resolution of conjunctions of equations
and disequations in the theory of finite or  infinite trees has been
studied by A. Colmerauer~\cite{col84} and H.J.
Bürckert~\cite{bur6}. An incremental algorithm for solving
conjunctions of equations and disequations on rational trees has
been proposed after by V.Ramachandran and P. Van
Hentenryck~\cite{ram42}. The resolution of universally quantified
disequations on finite trees has been also developed by A.
Smith~\cite{das}. We will find a general synthesis on this subject
in the work of H. Comon~\cite{com13}. M. Maher has
also shown that every formula is equivalent in $\cal T$ to a Boolean
combination of existentially quantified solved conjunctions of
elementary equations \cite{Maher}. Note that we get the same  result using Corollary \ref{thm}.

In what follows, we first show how to solve some simple formulas
without free variables in order to understand the application of
the rewriting rules and the role of each rule in ${\cal T}$,then
we give some benchmarks representing real situations on two
partner games by full first-order formulas with free variables.
\vspace{-4mm}
\subsection*{Simple examples}
\begin{example}
 Let us solve the following formula $\varphi_1$ in ${\cal T}$: \[\exists x\forall
y\,((\exists zwv\,y=fz\wedge y=fx\wedge w=gzv)\vee(x=fy \wedge
x=fx))\] Using Property  \ref{norm11} we first transform the
preceding formula into the following normalized formula
\begin{equation}\label{ex1} \neg(\exists\varepsilon\,\vrai\wedge\neg(\exists
x\,\vrai\wedge\neg\left[\begin{array}{l}\exists
y\,\vrai\wedge\\\neg(\exists zwv\,y=fz\wedge y=fx\wedge
w=gzv)\wedge\\\neg(\exists\varepsilon\,x=fy \wedge
x=fx)\end{array}\right]))\end{equation}Since $A=FL$ then the preceding  normalized formula is a working formula. Let us  decompose the
sub-formula
\begin{equation}\label{ex2}\exists zwv\,y=fz\wedge y=fx\wedge
w=gzv.\end{equation} According to Section \ref{dex}, the order
$\succ$ is chosen such that $z\succ w\succ v\succ y\succ x$. Using
the rewriting rules of Property \ref{resolution}, the sub-formula
$y=fz\wedge y=fx\wedge w=gzv$ is equivalent in ${\cal T}$ to the
$(\succ\!)$-solved formula $y=fz\wedge z=x\wedge w=gzv$, and thus
according to Section \ref{dex} the decomposed formula of
(\ref{ex2}) is
\[\exists z\,y=fz\wedge z=x\wedge(\exists v\,\vrai\wedge(\exists
w\,w=gzv))\] Since $(\exists
w\,w=gzv)\neq(\exists\varepsilon\,\vrai)$ we can apply the rule
(3) with $I=\emptyset$, thus the formula (\ref{ex1}) is
equivalent in ${\cal T}$ to
\begin{equation}\label{ex3} \neg(\exists\varepsilon\,\vrai\wedge\neg(\exists
x\,\vrai\wedge\neg\left[\begin{array}{l}\exists
y\,\vrai\wedge\\\neg(\exists zv\,y=fz\wedge
z=x)\wedge\\\neg(\exists\varepsilon\,x=fy \wedge
x=fx)\end{array}\right]))\end{equation} The sub-formula $\exists
zv\,y=fz\wedge z=x$ is not an element of $A'$ and is equivalent in
${\cal T}$ to the decomposed formula $\exists z\,y=fz\wedge
z=x\wedge(\exists v\,\vrai\wedge(\exists\varepsilon\,\vrai))$,
thus we can apply the rule (4) with $I=\emptyset$ and the formula
(\ref{ex3}) is equivalent in ${\cal T}$ to
\begin{equation}\label{ex4} \neg(\exists\varepsilon\,\vrai\wedge\neg(\exists
x\,\vrai\wedge\neg\left[\begin{array}{l}\exists
y\,\vrai\wedge\\\neg(\exists z\,y=fz\wedge
z=x)\wedge\\\neg(\exists\varepsilon\,x=fy \wedge
x=fx)\end{array}\right]))\end{equation} Let us decompose now the
sub-formula \begin{equation}\label{ex5}\exists\varepsilon\,x=fy
\wedge x=fx\end{equation} Using the rewriting rules of Property
\ref{resolution}, the sub-formula $x=fy\wedge x=fx$ is equivalent
in ${\cal T}$ to the $(\succ\!)$-solved formula $x=fy\wedge y=x$
and thus according to Section \ref{dex} the decomposed formula of
(\ref{ex5}) is
\[\exists\varepsilon\,x=fy\wedge
y=x\wedge(\exists\varepsilon\,\vrai\wedge(\exists\varepsilon\,\vrai))\]
Since $(\exists\varepsilon\,x=fy \wedge x=fx)\not\in A'$ then we
can apply the rule (4) with $I=\emptyset$ and thus the formula
(\ref{ex4}) is equivalent in ${\cal T}$ to
\begin{equation}\label{ex6} \neg(\exists\varepsilon\,\vrai\wedge\neg(\exists
x\,\vrai\wedge\neg\left[\begin{array}{l}\exists
y\vrai\wedge\\\neg(\exists z\,y=fz\wedge
z=x)\wedge\\\neg(\exists\varepsilon\,x=fy \wedge
y=x)\end{array}\right]))\end{equation} According to Section
\ref{dex} the formula $\exists\varepsilon\,\vrai\wedge(\exists
y\,\vrai\wedge(\exists\varepsilon\,\vrai))$ is the decomposed
formula of $\exists y\,\vrai$. Since $\exists y\,\vrai\not\in A'$,
$(\exists z\,y=fz\wedge z=x)\in A'$ and $(\exists\varepsilon\,x=fy
\wedge y=x)\in A'$ then we can apply the rule (4) and thus the
formula (\ref{ex6}) is equivalent in ${\cal T}$ to

\begin{equation}\label{ex7b} \neg(\exists\varepsilon\,\vrai\wedge\neg(\exists
\varepsilon\,\vrai\wedge\neg(\exists
\varepsilon\,\vrai))\end{equation} Finally, we can apply the rule
(1) thus the formula (\ref{ex7b}) is equivalent in ${\cal T}$ to
$
\neg(\exists\varepsilon\,\vrai)$. Thus $\varphi_1$ is false in
${\cal T}$.
\end{example}
\begin{example}
Let us solve the following  formula $\varphi_2$ in ${\cal T}$:
\begin{equation}\label{ex7} \exists
x\,\forall y\,((\exists z\,y=fz\wedge
z=x)\vee(\exists\varepsilon\,x=fy \wedge y=x)\vee\neg(x=fy))
\end{equation}
Using Property \ref{norm11} we first transform the preceding
formula into the following normalized formula

\begin{equation}\label{ex8} \neg(\exists\varepsilon\,\vrai\wedge\neg(\exists
x\,\vrai\wedge\neg\left[\begin{array}{l}\exists
y\,x=fy\wedge\\\neg(\exists z\,y=fz\wedge
z=x)\wedge\\\neg(\exists\varepsilon\,x=fy \wedge
y=x)\end{array}\right]))\end{equation} Since $A=FL$ then the preceding normalized formula is a working formula in ${\cal T}$. Since $(\exists y\,x=fy)\in
A'$, $(\exists z\,y=fz\wedge z=x)\in A'$ and
$(\exists\varepsilon\,x=fy \wedge y=x)\in A'$ then we can apply
the rule (5), thus the formula (\ref{ex8}) is equivalent in ${\cal
T}$ to
\begin{equation}\label{ex9} \neg\left[\begin{array}{l}\exists\varepsilon\,\vrai\wedge\\\neg(\exists x\,\vrai\wedge\neg(\exists y\,x=fy))\wedge\\\neg(\exists x_1y_1z\,
x_1=fy_1\wedge y_1=fz\wedge z=x_1)\wedge\\\neg(\exists x_2y_2
\,x_2=fy_2\wedge x_2=fy_2 \wedge
y_2=x_2)\end{array}\right]\end{equation} According to Section
\ref{dex} the formula $\exists\varepsilon\,\vrai\wedge(\exists
x\,\vrai\wedge(\exists\varepsilon\,\vrai))$ is the decomposed
formula of $\exists x\,\vrai$. Since $(\exists x\,\vrai)\not\in
A'$ and $(\exists y\,x=fy)\in A'$ then we can apply the rule (4)
and thus the formula (\ref{ex9}) is equivalent in ${\cal T}$ to
\begin{equation}\label{ex10} \neg\left[\begin{array}{l}\exists\varepsilon\,\vrai\wedge\\\neg(\exists \varepsilon\,\vrai)\wedge\\\neg(\exists x_1y_1z\,
x_1=fy_1\wedge y_1=fz\wedge z=x_1)\wedge\\\neg(\exists x_2y_2
\,x_2=fy_2\wedge x_2=fy_2 \wedge
y_2=x_2)\end{array}\right]\end{equation} Finally we can apply the
rule (1), thus the formula (\ref{ex10}) is equivalent in ${\cal
T}$ to $\vrai$. Thus $\varphi_2$ is true  in
${\cal T}$.

\end{example}

\subsection*{Benchmarks: Two partner games} Let $(V,E)$ be a
directed graph, with $V$ a set of vertices and $E\subseteq V\times
V$ a set of edges. The sets $V$ and $E$ may be empty and the
elements of $E$ are also called positions. We consider a
two-partner game which, given an initial position $x_0$,
consists, one after another, in choosing a position $x_1$ such
that $(x_0, x_1)\in E$, then a position $x_2$ such that $(x_1,
x_2)\in E$ and so on. The first one who cannot play any more has
lost and the other one has won. For example the two following
infinite graphs correspond to the two following games:

\includegraphics*[width=12cm]{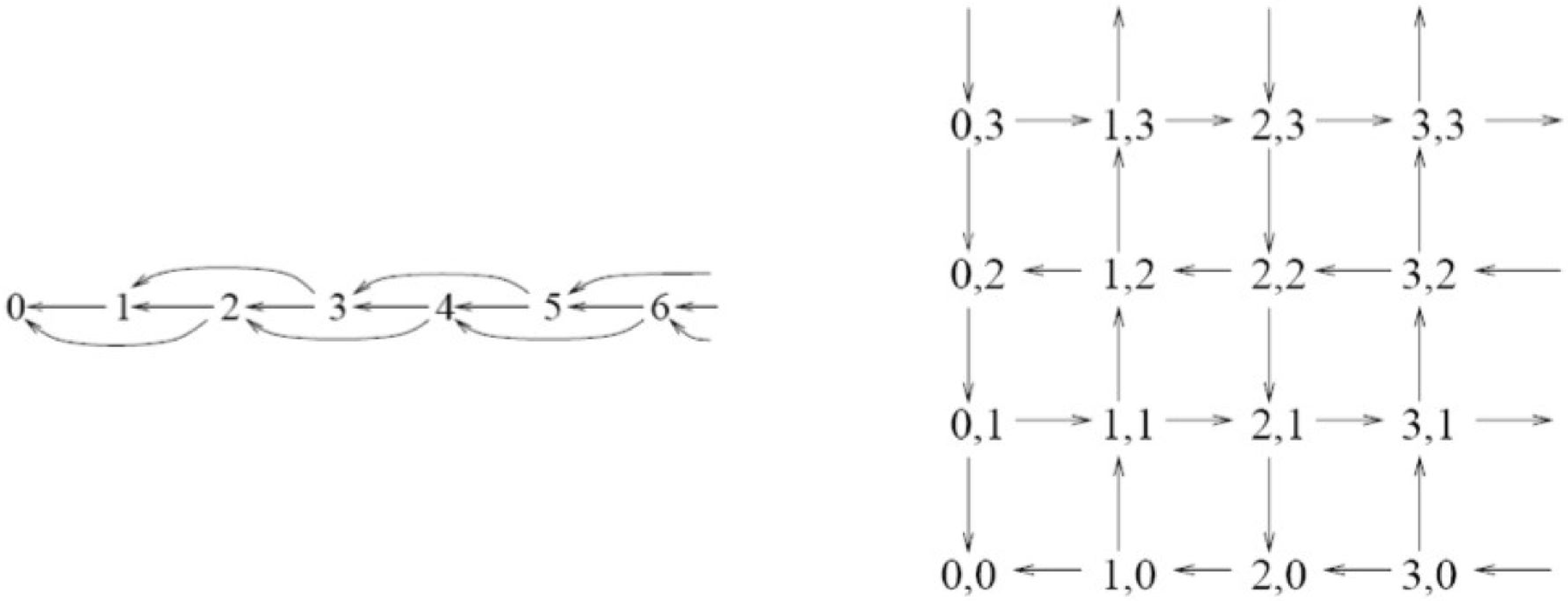}
$\\$
\begin{minipage}{4.5truecm} {\bfseries Game 1} A non-negative integer $i$ is given and, one after another,
each partner subtracts 1 or 2 from $i$, but keeping $i$
non-negative. The first person who cannot play any more has
lost.\end{minipage} \begin{minipage}{0.2truecm}\hspace{0mm}
\end{minipage}
\begin{minipage}{7.7truecm} {\bfseries Game 2} An ordered pair $(i,j)$ of non-negative
integers is given and, one after another, each partner chooses one
of the integers $i$, $j$. Depending on the fact that the chosen
integer $u$ is odd or even, he then increases or decreases the
other integer $v$ by 1, but keeping $v$ non-negative. The first
person who cannot play any more has lost.\end{minipage} $\\[1mm]$ Let
$x$ be a position in a game and suppose that it is the turn of
person A to play. The position $x$ is said to be $k$-winning if,
no matter the way the other person B plays, it is always possible
for A to win in making at most $k$ moves. It is easy to show that
\[winning_k(x)=\left[\begin{array}{l}\exists
y\,move(x,y)\wedge\neg(\\\exists
x\,move(y,x)\wedge\neg(\\...\\\exists
y\,move(x,y)\wedge\neg(\\\exists
x\,move(y,x)\wedge\neg(\\false\hspace{1.4cm}{\underbrace{)...)}}_{2k}\end{array}\right]\]
where move$(x,y)$ means : `` starting from the position $x$ we
play one time and reach the position $y$". By moving down the
negations, we get an embedding of 2k alternated quantifiers. We
represent this two games in the algebra of finite or  infinite trees
$(A,\Ff )$, where each position is represented by a tree.

If we take as input of our solver the formula $winning_k(x)$ we
will get as output a formula which represents all the $k$-winning
positions. $\\$ {\bfseries Game 1:} Suppose that $\Ff$ contains
the $0$-ary functional symbol $0$ and the $1$-ary functional
symbol $s$. We code the vertices $i$ of the game graph by the
trees $s^i(0)$%
\footnote{ Of course $s^0(x)=x$ and $s^{i+1}(x)=s(s^i(x))$.}
The relation $move(x,y)$ is defined as follows:
\[move(x,y)\stackrel{\mathrm{def}}{\leftrightarrow} x=s(y)\vee
x=s(s(y))\vee(\neg(x=0)\wedge\neg(\exists u\,x=s(u))\wedge x=y)\]
For $winning_1(x)$ our algorithm give the following solved
formula:
\[\neg\left[\begin{array}{l}\exists\varepsilon\,\vrai\wedge\left[\begin{array}{l}\neg(\exists u\,x=s(u) \wedge u=0)\wedge\\\neg(\exists
u_1u_2\, x=s(u_1)\wedge u_1=s(u_2) \wedge
u_2=0)\end{array}\right]\end{array}\right]\]which corresponds to
the solution $x = s(0) \vee x = s(s(0))$.$\\$ {\bfseries Game 2:}
Suppose that $\Ff$ contains the functional symbols $0$, $ f$, $g$,
$c$ of respective arities $0$, $1$, $1$, $2$. We code the vertices
$(i,j)$ of the game graph by the trees $c(\bar{i},\bar{j})$ with
$\bar{i} = (fg)^{i/2}(0)$ if $i$ is
even, and $\bar{i} = g(\overline{i-1})$ if $i$ is odd.%
\footnote{$(fg)^{0}(x)=x$ and $(fg)^{i+1}(x)=f(g({(fg)}^i(x)))$.}
The relation $move(x,y)$ is defined as follows:
\[move(x,y)\stackrel{\mathrm{def}}{\leftrightarrow}
transition(x,y)\vee (\neg(\exists uv\, x=c(u,v))\wedge x=y)\] with
\[\begin{array}{lll}
transition(x,y)&\stackrel{\mathrm{def}}{\leftrightarrow}&\left[\begin{array}{l}\exists
uvw\\\left[\begin{array}{l} (x=c(u,v)\wedge
y=c(u,w))\vee\\(x=c(v,u)\wedge
y=c(w,u))\end{array}\right]\\\wedge\\\left[\begin{array}{l}(\exists
i\,u=g(i)\wedge succ(v,w))\vee\\(\neg(\exists i\, u=g(i))\wedge
pred(v,w))\end{array}\right]\end{array}\right]
\\ \\
succ(v,w)&\stackrel{\mathrm{def}}{\leftrightarrow}&\left[\begin{array}{l}(\exists
j\, v=g(j)\wedge w=f(v))\vee\\(\neg(\exists j\,v=g(j)) \wedge
w=g(v))
\end{array}\right]
\end{array}\]
\[\begin{array}{lll}

pred(v,w)&\stackrel{\mathrm{def}}{\leftrightarrow}&\left[\begin{array}{l}(\exists
j\,v=f(j)\wedge\left[\begin{array}{l}(\exists k\,j=g(k)\wedge
w=j)\vee\\(\neg(\exists k\,j=g(k))\wedge
w=v)\end{array}\right])\vee\\(\exists j\,v=g(j)\wedge
\left[\begin{array}{l}(\exists k\,j=g(k)\wedge
w=v)\vee\\(\neg(\exists k\,j=g(k))\wedge
w=j)\end{array}\right])\vee\\
(\neg(\exists j\,v=f(j))\wedge\neg(\exists
j\,v=g(j))\wedge\neg(v=0)\wedge w=v)\end{array}\right]
\end{array}\]
For $winning_1(x)$ our algorithm give the following solved
formula:
\[\neg\left[\begin{array}{l}\exists\varepsilon\,\vrai\wedge\left[\begin{array}{l}\neg(\exists u_1u_2u_3\,x=c(u_1,u_2)\wedge u_1=g(u_3)\wedge u_2=0\wedge u_3=0)\wedge\\
\neg(\exists u_1u_2u_3\,x=c(u_1,u_2)\wedge u_2=g(u_3)\wedge
u_1=0\wedge u_3=0)\end{array}\right]\end{array}\right]\] which
corresponds to the solution $x=c(g(0),0) \vee x=c(0, g(0)).$

 The
times of execution (CPU time in milliseconds) of the formulas
$winning_k(x)$ are given in the following table as well as a comparison with those of \cite{moi3}. The algorithm was
programmed in C++ and the benchmarks are performed on a 2.5Ghz
Pentium IV processor, with 1024Mb of RAM. $\\[1mm]$
\begin{tabular}{|c||c|c|c|c|c|c|c|c|} \hline k (Game 1) & 0 & 1 & 2 & 4 &
10 & 20 & 40 & 80\\\hline Our alg & 0 & 0 & 5 & 11 & 178 & 2630 &
59430 & 2553746\\\hline \cite{moi3} & 0 & 0 & 5 & 10& 150 & 2130 &
45430 & 1920110 \\\hline\hline k (Game 2) & 0 & 1 & 2 & 4 &
10 & 20 & 40 & 80\\\hline Our alg & 0 & 79 & 209 & 508 & 3830 & 162393
&$ - $&$ - $\\\hline \cite{moi3} & 0 & 75 & 180 & 420& 3040 & 123025 &
$ - $ & $ - $\\\hline\end{tabular}

These benchmarks were first introduced  by  A. Colmerauer and T. Dao. in \cite{dao2} where the first results of  the algorithm of T. Dao \cite{dao1} were presented.  We used the same benchmarks in   a joint work with T. Dao \cite{moi3}  where we gave  a more efficient  algorithm for solving first-order formulas in finite or infinite trees with better performances. The algorithm \cite{moi3} uses two strategies: (1) a top-down propagation of constraints: where all the super-formulas are propagated to the sub-formulas, then locally solved and finally  restored and so on. (2) A bottom-up distribution of sub-formulas to decrease the depth of the formulas. The restorations of  constraints defined in the first point uses a particular property which holds only for  the theory of finite or infinite trees. This algorithm \cite{moi3} gives good performances and the first step enables us to obtain quickly the solved formulas without losing time with solving sub-formulas which contradict their super-formulas. On the other hand our general algorithm defined in this paper can not use these strategies since it handles general decomposable theories. The main idea is to decompose at each level a quantified conjunction of atomic formulas and to propagate only the third section $A'''$ into the sub-formulas (rule 3). Then, the rule (4) decreases the size of the conjunction of sub-formulas and eliminates some quantifiers. Finally, the rule (5) decreases the depth of the working formulas using distribution. This algorithm computes the $k$-winning positions with the same bounds of performances for the values of $k$ as those of  \cite{moi3} but takes 5{\%}-30{\%} more time to compute them. This is due to the specific treatments used in \cite{moi3}. Unfortunately, this rate (5{\%}-30{\%}) grows with the size of $k$ and thus with the size of the initial working formula. Anyway, it must be noted that we were able to compute the k-winning
positions of game 1 with k = 80, which corresponds to a formula 
involving an alternated embedding of more than 160 quantifiers with a non-specific algorithm for finite of infinite trees.

\section{Discussion and conclusion}
We defined in this paper a new class of theories that we call
\emph{decomposable theories} and showed their completeness using a sufficient condition for the completeness of first-order theories. Informally, a decomposable theory is  a theory where each quantified conjunction of atomic formulas  can be decomposed into  three embedded sequences of quantifications having
 particular properties, which can be expressed with the help
of  $\exists?$, 
$\exists^{\Psi(u)}_{\infty}$ and $\exists!$. We  deduced from this definition a sufficient condition so that a theory accepts full elimination of quantifiers and showed that there is a strong relation between the  set $A'$ and the notion of full elimination of quantifiers. We have also given a general algorithm for solving
 first-order formulas in any decomposable theory $T$.  This algorithm is
given in the form of a set of five rewriting which transform a working formula $\varphi$  to a wnfv conjunction $\phi$ of solved formulas. In particular if $\varphi$ is a proposition, then $\phi$ is either the formula $\vrai$ or $\neg\vrai$. 

On the other hand  S. Vorobyov~\cite{vo14} has shown that the
problem of deciding if a proposition is
true or not in the theory of finite or infinite trees is non-elementary, i.e. the
complexity of all algorithms solving propositions is not bounded by a
tower of powers of $2's$ (top down evaluation) with a fixed
height. A. Colmerauer and T. Dao \cite{dao2} have also given a
proof of non-elementary complexity of solving constraints in this
 theory.  As a consequence, the complexity of our
algorithm and the size of our solved formulas are of this order.
We can  easily show that the size of our solved formulas is bounded
above by a top down tower of powers of $2's$, whose height is the
maximal depth of nested negations in the initial formula. The
function $\alpha(\varphi)$ used to show the termination of our
rules illustrates this result. However, despite  this high
complexity, we have implemented our algorithm and solved some
benchmarks in ${\cal T}$ with formulas having long nested
alternated quantifiers (up to 160). This algorithm has given competitive results in term of maximal depth of formulas that can be solved, compared with those of \cite{moi3} but took more time to compute the solved formulas.  As a consequence, we are planning  with  Thom Fruehwirth \cite{thom}  to add to CHR a general mechanism to treat our normalized formulas. This will enable us to implement quickly and easily other versions of our algorithms in order to get  better  performances.

Currently, we are trying to find  a more abstract characterization  and/or a model theoretical  characterization of the decomposable theories. The current definition gives only an algorithmic insight into what it means for a theory to be complete. We expect to add new vectorial quantifiers in the decomposition such as $\exists^n$  which means \emph{there exists  $n$} and  $\exists^{\Psi(u)}_{0,\infty} $ which means   \emph{there exists zero or infinite}, in order to increase the size of the set of  decomposable theories and may be get a much more simple  definition  than the one defined in this paper. Another interesting challenge is to find which special quantifiers must be added to the decomposable theories to get an equivalence between complete theory and decomposable theory. A first attempt  on this subject is actually in progress using the quantifiers $\exists^n$ and  $\exists^{\Psi(u)}_{0,\infty}$. It would be also interesting to show if these new quantifiers are enough to prove that every theory which accepts elimination of quantifiers is decomposable. 

We have also established a long list of decomposable theories. We can cite for example: the theory of finite trees,  of infinite trees,  of finite or infinite trees \cite{moi3},  of
additive  rational or real numbers  with addition and subtraction,  of linear dense order  without endpoints, of ordered additive  rational or real numbers  with addition, subtraction and a linear dense order relation without endpoints, of the combination of tress and ordered additive rational numbers \cite{moi2}, of the construction of trees on an ordered set \cite{moi1}, of the extension into  trees of first-order theories  \cite{moi4}. It would also be interesting  to build some theories that can be decomposed using two completely different sets of $A$, $A'$, $A''$, $A'''$ and $\Psi(u)$ and  find syntactic or semantic relations between these sets. 

Currently, we are showing the decomposability of other fundamental theories such as:  theory of lists using a combination of particular trees, theory of queues as done in \cite{ryb}, and the combination of trees  and
real numbers together with addition, 
subtraction, multiplication and a linear dense 
order relation without endpoints. We are also trying  to find  some
formal methods  to get easily the sets $\psi(u)$, $A$, $A'$, $A''$ and $A'''$ for any decomposable theory $T$.

$\\[1mm]$
{\bfseries{Acknowledgments}} I thank Thi-Bich-Hanh Dao for our many discussions about the theory of finite or infinite trees. I thank her too for the quality of her remarks and advice on how to improve the organization of this paper. Particular  thanks also to  Alain Colmerauer for our
long discussions about decomposability and solving first-order constraints. I dedicate this paper  to him with my best
wishes for a fast recovery. Many thanks also to the anonymous referees for their careful reading and suggestions.


\begin{thebibliography}{}
\bibitem[\protect\citeauthoryear{Benhamou}{Benhamou}{1996}]{ben} Benhamou, F.,   Colmerauer, A.,  Garetta, H., Pasero, R. and Van-caneghem, M. 1996. Le manuel de Prolog IV. PrologIA, Marseille,
France.

\bibitem[\protect\citeauthoryear{Burckert}{Burkert}{1988}]{bur6}  Burckert, H. 1988. Solving disequations in equational theories. In Proceeding of the 9th Conference on
Automated Deduction, LNCS  310,  pp. 517--526, Springer-Verlag.


\bibitem[\protect\citeauthoryear{Clark}{Clark}{1978}]{clark} Clark, K.L. 1978. Negation as failure. In Logic and Data bases. Ed Gallaire, H. and Minker, J. Plenum Pub.
 

\bibitem[\protect\citeauthoryear{Colmerauer}{Colmerauer}{1982}]{col7} Colmerauer, A. 1982. Prolog and infinite trees. In K.L. Clark and S-A.
Tarnlund, editors, Logic Programming. Academic Press. pp. 231--251.

\bibitem[\protect\citeauthoryear{Colmerauer}{Colmerauer}{1984}]{col84}  Colmerauer, A. 1984. Equations and inequations on finite and infinite
trees. Proceeding of the International conference on the fifth
generation of computer systems,  pp. 85--99.


\bibitem[\protect\citeauthoryear{Colmerauer}{Colmerauer}{1990}]{Colmerauer90}  Colmerauer, A. 1990. An introduction to Prolog III. Communication of the ACM, 33(7):68--90.

\bibitem[\protect\citeauthoryear{Colmerauer}{Colmerauer}{2003}]{dao2}  Colmerauer, A. and Dao, T.  2003. Expressiveness of full first-order formulas  in the algebra of finite or infinite trees,
Constraints,  8(3):  283--302.

\bibitem[\protect\citeauthoryear{Comon}{Comon}{1988}]{com}  Comon, H. 1988. Unification et disunification : Theorie et applications. PhD thesis, Institut National Polytechnique de
Grenoble.

\bibitem[\protect\citeauthoryear{Comon}{Comon}{1989}]{com15}  Comon, H. and
 Lescanne, P. 1989. Equational problems and disunification.
Journal of Symbolic Computation, 7: 371--425.

\bibitem[\protect\citeauthoryear{Comon}{Comon}{1991}]{com13} Comon, H. 1991. Disunification: a survey.
In J.L. Lassez and G. Plotkin, editors, Computational Logic:
Essays in Honor of Alan Robinson. MIT Press.
\bibitem[\protect\citeauthoryear{Comon}{Comon}{1991}]{com14}  Comon, H. 1991. Resolution de contraintes
dans des algebres de termes. Rapport d'Habilitation, Universite de
Paris Sud.
\bibitem[\protect\citeauthoryear{Courcelle}{Courcelle}{1983}]{cou1}  Courcelle, B. 1983. Fundamental Properties of Infinite Trees, Theoretical Computer Science,  25(2):95--169.

\bibitem[\protect\citeauthoryear{Courcelle}{Courcelle}{1986}]{cou2} Courcelle, B. 1986. Equivalences and Transformations of Regular
Systems applications to Program Schemes and Grammars, Theoretical
Computer Science, 42: 100--122.


\bibitem[\protect\citeauthoryear{Dao}{Dao}{2000}]{dao1} Dao, T. 2000. Resolution de contraintes du premier ordre dans la theorie des arbres finis ou infinis. These d'informatique, Universite de la mediterranee, France.



\bibitem[\protect\citeauthoryear{Djelloul}{Djelloul}{2005a}]{moi1} Djelloul, K. 2005a. Complete first-order axiomatization of the construction of trees on an ordered
set. Proceedings of the 2005 International Conference on
Foundations of Computer Science (FCS'05),  CSREA Press, pp. 87--93.

\bibitem[\protect\citeauthoryear{Djelloul}{Djelloul}{2005b}]{moi2} 
Djelloul, K. 2005b. About the combination of trees and rational
numbers in a complete first-order theory. Proceeding of the 5th
International conference on frontiers of combining systems FroCoS
2005,  Springer Lecture Notes in Artificial
Intelligence, vol 3717, pp. 106--122.

\bibitem[\protect\citeauthoryear{Djelloul}{Djelloul}{2006a}]{moi3} 
Djelloul, K. and Dao, T. 2006a.  Solving First-Order formulas  in the Theory of Finite or Infinite Trees : Introduction to the Decomposable Theories. Proceeding of the 21st ACM Symposium on Applied Computing (SAC). ACM press (to appear).

\bibitem[\protect\citeauthoryear{Djelloul}{Djelloul}{2006b}]{moi4} 
Djelloul, K. and Dao, T. 2006b.   Complete first-order axiomatization of the M-extended trees. Proceeding of the 20th  Workshop on (constraint) Logic Programming (WLP06). INFSYS Research Report 1843-06-02, pp. 111--119.

\bibitem[\protect\citeauthoryear{Fruehwirth}{Fruehwirth}{2002}]{thom} Fruehwirth T., Abdelnnadher S. Essentials of constraints programming. Springer Cognitive technologies. 


\bibitem[\protect\citeauthoryear{Huet}{Huet}{1976}]{hue}  Huet, G. 1976. Resolution d'equations dans les langages
d'ordre 1, 2,\ldots $\omega$. These d'Etat, Universite Paris 7.
France.

\bibitem[\protect\citeauthoryear{Jaffar}{Jaffar}{1984}]{jaf28} Jaffar, J. 1984. Efficient unification over infinite terms. New Generation Computing,
2(3): 207--219.
\bibitem[\protect\citeauthoryear{Jhon}{Jhon}{1979}]{j26}  John, E. and  Ullman, D. 1979. Introduction to automata theory,
languages and computation. Addison-Wesley publishing company.
\bibitem[\protect\citeauthoryear{Kunen}{Kunen}{1987}]{kun31}  Kunen, K. 1987. Negation in logic programming. Journal of Logic Programming, 4: 289--308.
\bibitem[\protect\citeauthoryear{Lyndo}{Lyndon}{1964}]{lyn}  Lyndon, R.C. 1964. Notes on logic. Van Nostrand
Mathematical studies.
\bibitem[\protect\citeauthoryear{Maher}{Maher}{1988}]{Maher}  Maher, M. 1988. Complete axiomatization of the algebra of finite, rational and infinite trees. Technical report, IBM - T.J.Watson Research
Center.
\bibitem[\protect\citeauthoryear{Malcev}{Malcev}{1971}]{mal38}  Malcev, A. 1971. Axiomatizable classes of locally free algebras of various types. In B.Wells
III, editor, The Metamathematics of Algebraic Systems. Anatolii
Ivanovic Malcev. Collected Papers: 1936-1967, volume 66, chapter
23, pp. 262--281.
\bibitem[\protect\citeauthoryear{Matelli}{Matelli}{1982}]{mat39}  Matelli, A. and  Montanari, U. 1982. An efficient unification
algorithm. ACM Trans. on Languages and Systems, 4(2): 258--282.
\bibitem[\protect\citeauthoryear{Paterson}{Paterson}{1978}]{pat41} Paterson, M. and  Wegman, N. 1978. Linear unification. Journal of Computer and Systems Science,
16:158--167.



\bibitem[\protect\citeauthoryear{Ramachandran}{Ramachandran}{1993}]{ram42}
Ramachandran, V. and  Van Hentenryck, P. 1993. Incremental
algorithms for formula   solving and entailment over rational
trees. Proceeding of the 13th Conference Foundations of Software
Technology and Theoretical Computer Science, LNCS volume 761, pp.  205--217.
\bibitem[\protect\citeauthoryear{Robinson}{Robinson}{1965}]{rob43}  Robinson, J.A. 1965. A machine-oriented logic based on the
resolution principle. JACM, 12(1):23--41.

\bibitem[\protect\citeauthoryear{Rybina}{Rybina}{2001}]{ryb} Rybina, T. and Voronkov, A. 2001. A decision procedure for term algebras with queues. ACM transaction on computational logic. 2(2): 155-181.

\bibitem[\protect\citeauthoryear{Smith}{Smith}{1991}]{das}
 Smith, A. 1991. Constraint   operations for CLP. In Logic
Programming: Proceedings of the 8th International Conference.
Paris. pp. 760--774.


\bibitem[\protect\citeauthoryear{Vorobyov}{Vorobyov}{1996}]{vo14}  Vorobyov, S. 1996. An Improved Lower Bound for the Elementary Theories of
Trees, Proceeding of the 13th International Conference on
Automated Deduction (CADE'96). Springer Lecture Notes in
Artificial Intelligence, vol 1104, pp. 275-- 287.
\end{thebibliography}
\end{document}